\RequirePackage{ifpdf}
\documentclass[12pt,letterpaper]{article}
\pdfoutput=1
\usepackage{jheppub}
\usepackage{epsfig}
\usepackage{enumitem}
\usepackage{bbm}
\usepackage{rotating,graphicx}
\usepackage{amssymb,amsmath,amsfonts,mathtools}
\usepackage{fancybox,diagbox}
\usepackage{enumerate}
\usepackage{dsfont}
\usepackage{verbatim}
\usepackage{wrapfig}
\usepackage{slashed}
\usepackage{shuffle}
\usepackage{braket}
\usepackage{pdflscape}
\usepackage{accents}
\usepackage{afterpage}
\usepackage[vcentermath]{youngtab}

\author{Marco S. Bianchi}
 
\affiliation[a]{Facultad de Ingeniería, Universidad San Sebastián, Santiago, Chile}

\emailAdd{marco.bianchi@uss.cl}
  
\title{Universal and Transcendental Structures in Protected ABJM Two-Point Functions}

\abstract{
We consider two-loop corrections to two-point functions of protected scalar operators in ABJM theory. We infer a compact finite-rank formula valid for operators of arbitrary dimension and multi-trace structure. The result is governed by the exact tree-level metric and a simple kernel on the Young lattice, while the planar limit reduces to a simple partition-theoretic rule. The required integrals organize into uniformly transcendental combinations, providing evidence for uniform transcendentality of these protected two-point functions.
}

\def\Tr{\textrm{Tr}}

\newcommand{\masterintegralpicN}[1]{\,\raisebox{\dimexpr\fontdimen22\textfont2-.5\height\relax}{\includegraphics[scale=0.1]{pictures/N#1.png}}}

\newcommand{\threeLoopIntegralPic}[1]{\,\raisebox{\dimexpr\fontdimen22\textfont2-.5\height\relax}{\includegraphics[scale=0.1]{pictures/#1.png}}}

\numberwithin{equation}{section}

\newlength{\dhatheight}

\begin{document}

\maketitle
\allowdisplaybreaks

\section{Introduction}

ABJM theory provides a highly constrained setting in which to study perturbative and non-perturbative aspects of three-dimensional quantum field theory. The model is an $\mathcal{N}=6$ superconformal Chern--Simons--matter theory with gauge group $U(N)_k\times U(N)_{-k}$, while its ABJ generalization allows unequal ranks $N_1$ and $N_2$ \cite{Aharony:2008ug,Aharony:2008gk}. Protected scalar operators are especially useful probes: their conformal dimensions are fixed by supersymmetry, but, unlike $\mathcal{N}=4$ SYM \cite{Lee:1998bxa,Intriligator:1999ff,Eden:1999gh,Arutyunov:2001qw,Heslop:2001gp,Baggio:2012rr}, their two-point-function normalizations remain nontrivial and contain detailed information about color contractions, operator mixing, and loop effects \cite{Bianchi:2011correlators,Young:2013formfactors,Young:2014lka,Young:2014sia,Bianchi:2020cfn,Bianchi:2024nah}.

Uniform transcendentality is one of the most striking analytic regularities
of perturbative $\mathcal{N}=4$ SYM. An early manifestation is the principle of maximal transcendentality, developed in studies
of BFKL/DGLAP evolution and anomalous dimensions
\cite{Kotikov:2002ab,Kotikov:2003fb,Kotikov:2004er}; it subsequently became
closely intertwined with integrability in the planar spectral problem
\cite{Eden:2006rx,Beisert:2006ez,Beisert:2010jr}. Uniform transcendentality
has also been observed extensively in scattering amplitudes
\cite{Bern:2005iz,Goncharov:2010jf}, light-like Wilson loops
\cite{DelDuca:2009au,DelDuca:2010zg}, and form factors of local operators
\cite{vanNeerven:1985ja,Bork:2010wf,Gehrmann:2011xn,Brandhuber:2012vm,Brandhuber:2014ica,Banerjee:2016kri,Huber:2019fxe,Lin:2020dyj,Agarwal:2021zft,Lee:2021lkc}.
After a suitable normalization, the coefficient of each power of the
dimensional regulator has a definite transcendental weight. From the
Feynman-integral perspective, this structure is closely related to the
existence of uniformly transcendental bases of master integrals and to
canonical systems of differential equations \cite{Henn:2013pwa}.

Closer to the present problem, dimensionally regulated two-point functions of protected operators in $\mathcal{N}=4$ SYM were found to exhibit uniform transcendentality to high orders in the regulator, both for the dimension-two operator and for higher-dimensional multi-trace operators \cite{Bianchi:2023llc,Bianchi:2026tracing}. These results suggest that protected correlators can serve not only as observables of intrinsic interest, but also as sensitive probes of the transcendental structure of the underlying propagator integrals.

Related transcendental structures have appeared in ABJM scattering amplitudes,
light-like Wilson loops, and form factors
\cite{Henn:2010abjmWilson,Chen:2011abjmDualities,Bianchi:2011amplitudeWilson,Bianchi:2011allorder,CaronHuot:2012abjmSixPoint,Bianchi:2013finiteN,Bianchi:2013nonplanar,Bianchi:2014threeLoopABJM,Bianchi:2013lightlike,Brandhuber:2013sudakov,Young:2013formfactors,Lee:2010yangian,Huang:2013orthogonal,Huang:2014loopGrassmannian}.
For the dimension-one protected correlator, the higher-order $\epsilon$
expansion likewise displays uniform transcendentality and motivates a
corresponding basis of three-loop propagator integrals \cite{Bianchi:2024nah}.
The four-loop propagator integrals required at higher operator dimension
therefore provide a natural testing ground for extending this pattern
\cite{Bianchi:2025sjc}.

In this work we compute two-loop two-point functions of protected scalar operators with general trace structure. We keep the full dependence on the ABJ ranks $N_1$ and $N_2$, use dimensional reduction in $d=3-2\epsilon$, and reduce the momentum-space integrals through integration-by-parts identities \cite{Tkachov:1981wb,Chetyrkin:1981qh,Laporta:2001dd}. Complete correlator matrices through dimension four are used to infer the general formula, which is then tested independently by a direct calculation of the full dimension-five matrix.

Our results exhibit two complementary simplifications. First, the required four-loop master integrals can be organized into a uniformly transcendental basis. One element of this basis is a nontrivial linear combination whose normalization is inferred directly from the uniform-transcendentality properties of a protected correlator. Second, all explicit higher-dimensional results depend on only three universal combinations of these master integrals. In the planar limit, the dependence on a multi-trace operator is controlled by the centralizer factor of its associated partition and by the number of length-two traces. At finite rank, the complete result is described by the exact tree-level metric together with a kernel induced by one-box steps on the Young lattice.

The paper is organized as follows. Section~\ref{sec:definitions} introduces the protected operators and perturbative conventions. Section~\ref{sec:tree} reviews the exact tree-level correlator matrix. Section~\ref{sec:two-loop} presents the two-loop calculation and the uniformly transcendental master-integral basis. Section~\ref{sec:full-planar} describes the planar all-length pattern, and Section~\ref{sec:finiteN} gives its finite-rank extension. We conclude with a discussion of open directions. The master-integral expansions used in the calculation are collected in Appendix~\ref{app:expansions}.

\section{Definitions}
\label{sec:definitions}

We consider ABJM theory and, to keep the color dependence explicit, retain its ABJ generalization with gauge group
\begin{equation}
U(N_1)_k\times U(N_2)_{-k}\,.
\end{equation}
The ABJM specialization is obtained by setting $N_1=N_2=N$. The model is an $\mathcal{N}=6$ superconformal Chern--Simons--matter theory in three dimensions, and perturbation theory is an expansion at large Chern--Simons level $k$ \cite{Aharony:2008ug,Aharony:2008gk}. We regulate loop integrals by dimensional reduction \cite{Siegel:1979wq} in
\begin{equation}
d=3-2\epsilon\,.
\end{equation}
This prescription keeps the three-dimensional field content and the algebra of the Levi-Civita tensors while analytically continuing loop momenta to $d$ dimensions. It is particularly convenient for supersymmetric observables in Chern--Simons--matter theories and for studying their transcendentality properties \cite{Bianchi:2024nah}.
In four-dimensional $\mathcal{N}=1,2,4$ super Yang--Mills theories,
dimensional reduction has recently been confirmed to preserve supersymmetry
through three loops, with earlier apparent violations traced to subtleties in
the treatment of the Clifford algebra \cite{Chakraborty:2026hdv}.
Motivated by these results and by its standard use in ABJM perturbation
theory, we assume throughout that dimensional reduction provides the
appropriate supersymmetry-preserving regularization scheme for the present
calculation.

The matter sector contains four complex scalars $Y^A$, with $A=1,\ldots,4$, transforming in the $\mathbf{4}$ of the $SU(4)$ R-symmetry and in the bifundamental representation $(\mathbf{N}_1,\overline{\mathbf{N}}_2)$ of the gauge group. Their conjugates $\bar Y_A$ transform in $(\overline{\mathbf{N}}_1,\mathbf{N}_2)$. We write their gauge indices as
\begin{equation}
(Y^A)^i{}_{\hat\jmath}\,,
\qquad
(\bar Y_A)^{\hat\imath}{}_{j}\,,
\end{equation}
where unhatted and hatted indices belong to $U(N_1)$ and $U(N_2)$, respectively. With canonical normalization, the free position-space propagator is
\begin{align}
\left\langle
(Y^A)^i{}_{\hat\jmath}(x)
(\bar Y_B)^{\hat l}{}_{m}(0)
\right\rangle_0
&=
\delta^A_B\,\delta^i_m\,\delta^{\hat l}_{\hat\jmath}\,
\Pi(x,\epsilon)\,,
\\
\Pi(x,\epsilon)
&\equiv
\frac{\Gamma\left(\frac12-\epsilon\right)}{4\pi^{\frac32-\epsilon}}
\frac{1}{\left(x^2\right)^{\frac12-\epsilon}}\,.
\label{eq:abjm-propagator}
\end{align}

The protected scalar operators studied below are built from alternating products of $Y$ and $\bar Y$. To select a definite highest-weight component of the half-BPS multiplet, we use the off-diagonal flavor combination $Y^1\bar Y_2$; any other choice related to it by $SU(4)$ is equivalent.

We now fix the convention used to label trace structures and correlator-matrix entries throughout the paper. At protected dimension $n$, let $p(n)$ be the number of partitions of $n$, and order them in decreasing lexicographic order,
\begin{equation}
\mathfrak{P}_n
=
\left\{\mu^{(1)},\ldots,\mu^{(p(n))}\right\}\,,
\qquad
\mu^{(i)}\,\vdash\,n\,.
\label{eq:ordered-partitions}
\end{equation}
Thus, $\mu^{(i)}$ precedes $\mu^{(j)}$ if, at the first position at which they differ, the corresponding part of $\mu^{(i)}$ is larger; missing parts are understood to be zero. The associated partition of the elementary-field length is
\begin{equation}
\lambda^{(i)}
\equiv
2\mu^{(i)}
=
\left(2\mu^{(i)}_1,\ldots,2\mu^{(i)}_{\ell_i}\right)\,.
\label{eq:even-partition}
\end{equation}
For each ordered partition we define
\begin{equation}
\mathcal{O}_{i}(x)
\equiv
\prod_{a=1}^{\ell_i}
\Tr\!\left[\left(Y^1\bar Y_2\right)^{\mu^{(i)}_a}\right](x)\,,
\qquad
\overline{\mathcal{O}}_{i}(x)
\equiv
\prod_{a=1}^{\ell_i}
\Tr\!\left[\left(Y^2\bar Y_1\right)^{\mu^{(i)}_a}\right](x)\,,
\qquad
\ell_i\equiv\ell\!\left(\mu^{(i)}\right)\,.
\label{eq:operators}
\end{equation}
Here the traces are taken over $U(N_1)$ gauge indices; the equivalent $U(N_2)$ representation follows by cyclically interchanging $Y$ and $\bar Y$. Each part $\mu^{(i)}_a$ labels a trace containing $\mu^{(i)}_a$ pairs $Y\bar Y$, or equivalently $2\mu^{(i)}_a$ elementary fields. Since an elementary scalar has classical dimension $1/2$, these operators have elementary-field length and protected conformal dimension
\begin{equation}
L=2n\,,
\qquad
\Delta=n\,.
\end{equation}
For example, at dimensions two, three, and four the ordered sets are
\begin{equation}
\begin{aligned}
n=2:\quad
&\left(\mu^{(1)},\mu^{(2)}\right)
=\left((2),(1,1)\right)\,,
\qquad
\left(\lambda^{(1)},\lambda^{(2)}\right)
=\left((4),(2,2)\right)\,,
\\[0.3em]
n=3:\quad
&\left(\mu^{(1)},\mu^{(2)},\mu^{(3)}\right)
=\left((3),(2,1),(1,1,1)\right)\,,
\\
&\left(\lambda^{(1)},\lambda^{(2)},\lambda^{(3)}\right)
=\left((6),(4,2),(2,2,2)\right)\,,
\\[0.3em]
n=4:\quad
&\left(\mu^{(1)},\ldots,\mu^{(5)}\right)
=\left((4),(3,1),(2,2),(2,1,1),(1,1,1,1)\right)\,,
\\
&\left(\lambda^{(1)},\ldots,\lambda^{(5)}\right)
=\left((8),(6,2),(4,4),(4,2,2),(2,2,2,2)\right)\,.
\end{aligned}
\label{eq:partition-order-examples}
\end{equation}
Accordingly, $G_{3;1,2}$ denotes the dimension-three correlator between the single-trace operator of length six and the double-trace operator with trace lengths $(4,2)$. More generally, the index preceding the semicolon in $G_{n;i,j}$ specifies the protected dimension, while $i$ and $j$ identify the ordered partitions in $\mathfrak{P}_n$. This convention fixes unambiguously all matrix entries displayed below.

Unlike in four-dimensional $\mathcal{N}=4$ SYM, protection fixes the conformal dimension but does not make the two-point-function normalization tree-level exact \cite{Bianchi:2024nah}. At finite $N_1$ and $N_2$, operators associated with different partitions can mix, so we retain the full correlator matrix in the trace basis rather than assuming orthogonality.

At tree level, a correlator of operators labelled by $i,j\in\{1,\ldots,p(n)\}$ contains $2n$ scalar propagators and can be written as
\begin{equation}
\left\langle
\mathcal{O}_{i}(x)
\overline{\mathcal{O}}_{j}(0)
\right\rangle_0
=
\Pi(x,\epsilon)^{2n}\,
G^{(0)}_{n;i,j}(N_1,N_2)\,,
\label{eq:tree}
\end{equation}
where the exact finite-rank color matrix $G^{(0)}_{n;i,j}$ is described in the next section. Odd-loop corrections to these scalar two-point functions vanish \cite{Bianchi:2024nah}, and we organize the perturbative expansion as
\begin{equation}
\left\langle
\mathcal{O}_{i}(x)
\overline{\mathcal{O}}_{j}(0)
\right\rangle
=
\Pi(x,\epsilon)^{2n}
\left[
G^{(0)}_{n;i,j}
+\frac{1}{k^2}\,
\left(16\pi e^{\gamma_E}x^2\right)^{2\epsilon}
G^{(2)}_{n;i,j}
+\mathcal{O}\!\left(k^{-4}\right)
\right]\,.
\label{eq:perturbative-expansion}
\end{equation}
The extra factor $\left(16\pi e^{\gamma_E}x^2\right)^{2\epsilon}$ has been factored out of $G^{(2)}_{n;i,j}$ to keep the expressions presented below cleaner.

\section{Tree-level correlators}
\label{sec:tree}

The finite-rank tree-level correlators of the operators considered here are known from the Schur-polynomial description of the half-BPS sector of ABJM theory \cite{Dey:2011ea,Chakrabortty:2011gv,Caputa:2012pi}. The analogous Schur-polynomial organization of exact finite-rank half-BPS correlators was established earlier in $\mathcal{N}=4$ SYM \cite{Corley:2001zk}, with extensions to multi-matrix operators and restricted Schur bases developed in \cite{Brown:2007multimatrix,Bhattacharyya:2008multimatrix}. We briefly recast the ABJM results in a form that will be useful for the analysis of the loop corrections.

For operators of length
\begin{equation}
L=2n\,,
\end{equation}
the trace-basis operators are labelled by the ordered partitions $\mu^{(i)}\in\mathfrak{P}_n$ introduced in \eqref{eq:ordered-partitions}. Let $R\,\vdash\,n$ denote a Young diagram, and let $\chi_R(\mu^{(i)})$ be the character of the symmetric group $S_n$ in the irreducible representation $R$, evaluated on the conjugacy class of cycle type $\mu^{(i)}$.

For a box $(a,b)\in R$, let $c(a,b)=b-a$ denote its content. The
standard content product associated with $R$ (equivalently, the generalized
Pochhammer symbol for the partition $R$) is
\begin{equation}
f_R(N)
=
\prod_{(a,b)\in R}
\left(N+b-a\right)\,,
\label{eq:fR-content-product}
\end{equation}
By the hook-content and hook-length formulae, it may equivalently be written
as
\begin{equation}
f_R(N)
=
\frac{n!\,\dim_{U(N)}R}{d_R}\,,
\end{equation}
where $d_R$ is the dimension of the irreducible representation $R$ of
$S_n$. We then define the bifundamental Schur-basis norm
\begin{equation}
F_R(N_1,N_2)
=
f_R(N_1)f_R(N_2)\,.
\label{eq:FRdef}
\end{equation}
Denoting the Schur-basis operators by $s_R$, their tree-level two-point function is diagonal,
\begin{equation}
\left\langle
s_R\,\overline{s}_S
\right\rangle_0
=
\delta_{RS}\,
F_R(N_1,N_2)\,.
\label{eq:tree-schur}
\end{equation}

Using the Frobenius relation
\begin{equation}
p_{\mu^{(i)}}
=
\sum_{R\,\vdash\,n}
\chi_R\!\left(\mu^{(i)}\right)s_R\,,
\end{equation}
the exact tree-level correlator in the multi-trace basis can be written as
\begin{equation}
G^{(0)}_{n;i,j}
=
\sum_{R\,\vdash\,n}
\chi_R\!\left(\mu^{(i)}\right)
\chi_R\!\left(\mu^{(j)}\right)
F_R(N_1,N_2)\,.
\label{eq:tree-trace}
\end{equation}
This form separates completely the combinatorial dependence on the trace structures, encoded in the symmetric-group characters, from the dependence on the gauge-group ranks, contained in $F_R(N_1,N_2)$.

For later convenience, the correlators at fixed $n$ may also be encoded in the generating function
\begin{equation}
\mathcal{G}_n(\mathbf{p},\mathbf{q})
=
\sum_{i,j=1}^{p(n)}
\frac{
G^{(0)}_{n;i,j}
}{
z_{\mu^{(i)}}z_{\mu^{(j)}}
}
p_{\mu^{(i)}}q_{\mu^{(j)}}\,,
\end{equation}
where $m_r(\mu^{(i)})$ denotes the multiplicity of the part $r$ in $\mu^{(i)}$. The associated centralizer factor is
\begin{equation}
z_{\mu^{(i)}}
=
\prod_{r\geq 1}
r^{m_r(\mu^{(i)})}
\bigl(m_r(\mu^{(i)})\bigr)!\,.
\label{eq:zmu}
\end{equation}
Using \eqref{eq:tree-trace}, this becomes
\begin{equation}
\mathcal{G}_n(\mathbf{p},\mathbf{q})
=
\sum_{R\,\vdash\,n}
F_R(N_1,N_2)\,
s_R(\mathbf{p})\,
s_R(\mathbf{q})\,.
\label{eq:tree-generating}
\end{equation}
The corresponding all-length generating function is
\begin{equation}
\mathcal{G}(t;\mathbf{p},\mathbf{q})
=
\sum_R
t^{|R|}
F_R(N_1,N_2)\,
s_R(\mathbf{p})\,
s_R(\mathbf{q})\,.
\label{eq:tree-generating-all}
\end{equation}

The planar limit follows immediately from \eqref{eq:tree-trace}. Taking $N_1,N_2\sim N\to\infty$ at fixed ratio, we have
\begin{equation}
F_R(N_1,N_2)
=
\left(N_1N_2\right)^n
+
\mathcal{O}\left(N^{2n-1}\right)\,,
\end{equation}
orthogonality of the symmetric-group characters gives
\begin{equation}
G^{(0)}_{n;i,j}
=
z_{\mu^{(i)}}
\left(N_1N_2\right)^n
\delta_{ij}
+
\text{subleading terms in }N_1,N_2\,.
\label{eq:tree-planar}
\end{equation}
Thus, at leading order in the ranks, the multi-trace basis is orthogonal and the norm of its $i$th element is controlled by the standard combinatorial factor $z_{\mu^{(i)}}$. The same factor will reappear naturally in the planar part of the two-loop correction.

\section{Two-loop corrections}\label{sec:two-loop}

We begin by reviewing the dimension-one two-point function, which fixes our normalization and illustrates uniform transcendentality in its simplest form. We then describe the extension to higher-dimensional operators and extract patterns that can be extrapolated to arbitrary length.

\subsection{Dimension-one correlator}

The two-loop dimension-one correlator was revisited and its higher-order $\epsilon$ expansion was shown to exhibit uniform transcendentality \cite{Bianchi:2024nah}. This observation motivates the following uniformly transcendental basis of three-loop propagator integrals \cite{Bianchi:2025sjc}:
\begin{equation}
\begin{gathered}
\begin{aligned}
U_1^{(3)}
&\equiv
\frac{(1-4\epsilon)(1-6\epsilon)}{\epsilon^2}
\threeLoopIntegralPic{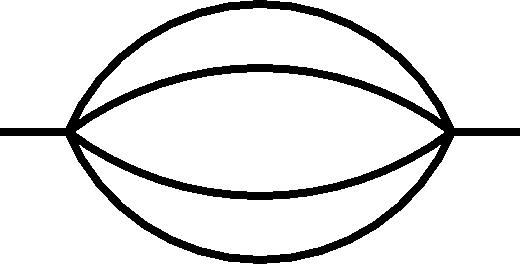}\,,
&\qquad
U_2^{(3)}
&\equiv
\frac{1-6\epsilon}{\epsilon}
\threeLoopIntegralPic{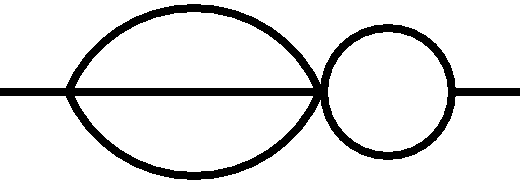}\,,
\\[0.5em]
U_3^{(3)}
&\equiv
\threeLoopIntegralPic{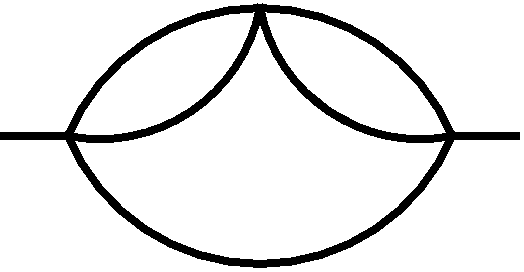}\,,
&
U_4^{(3)}
&\equiv
\threeLoopIntegralPic{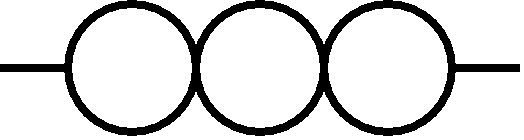}\,.
\end{aligned}
\\[0.5em]
U_5^{(3)}
\equiv
\threeLoopIntegralPic{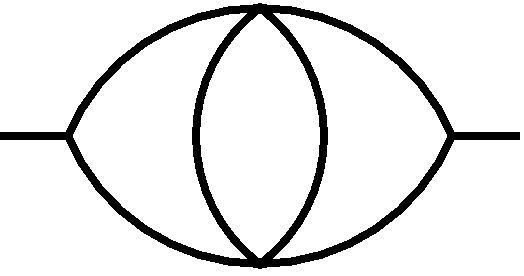}\,.
\end{gathered}
\label{eq:three-loop-ut-first-five}
\end{equation}
The final element is the following linear combination of three-loop integrals:
\begin{align}
U_6^{(3)}
&\equiv
\frac{1+2\epsilon}{4(1+4\epsilon)}
\threeLoopIntegralPic{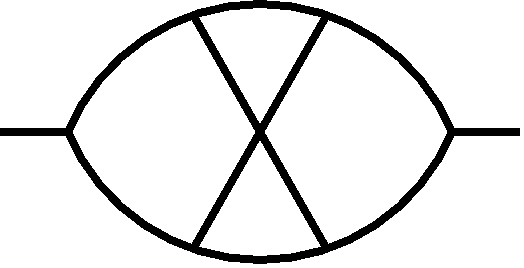}
+
\frac{5\epsilon(1+6\epsilon)}{(1+2\epsilon)(1+4\epsilon)}
\threeLoopIntegralPic{M3L5}
\nonumber\\
&\quad
+
\frac{192\epsilon^2}{(1+2\epsilon)^2}
\threeLoopIntegralPic{M3L3}
+
\frac{14\epsilon(1-6\epsilon)(1+6\epsilon)}{(1+2\epsilon)^2(1+4\epsilon)}
\threeLoopIntegralPic{M3L2}
\nonumber\\
&\quad
-
\frac{2(1-4\epsilon)(1-6\epsilon)
\left(172\epsilon^2+60\epsilon+3\right)}
{(1+2\epsilon)^3(1+4\epsilon)}
\threeLoopIntegralPic{M3L1}\,.
\label{eq:three-loop-ut-six}
\end{align}

For $U_1^{(3)},\ldots,U_5^{(3)}$, uniform transcendentality follows from their analytic representations in terms of gamma and hypergeometric functions. For $U_6^{(3)}$, it has been verified through transcendental weight ten and is assumed to persist to all orders in the $\epsilon$ expansion \cite{Bianchi:2024nah}.

At dimension one, $\mathfrak{P}_1=\{(1)\}$, so the correlator matrix has a
single entry. Fourier transformation to position space and normalization by
the two free propagators in \eqref{eq:perturbative-expansion} produce the
overall factor
\begin{equation}
\mathcal{Z}_{\mathrm{FT}}^{(3)}(\epsilon)
\equiv
\frac{
4^{\frac52-3\epsilon}\,
e^{-5\gamma_E\epsilon}\,\pi^{\frac72}\,
\Gamma(1-4\epsilon)
}{
\Gamma\!\left(\frac12-\epsilon\right)^2
\Gamma\!\left(\frac12+3\epsilon\right)
}\,.
\label{eq:dimension-one-FT-factor}
\end{equation}
Its gamma-function representation makes its uniform transcendentality
manifest, so it preserves the uniform weight of the master-integral
combinations below.
In the conventions of \eqref{eq:perturbative-expansion}, the two-loop
coefficient is therefore\footnote{Equations (3.18) and (4.9) of the published version of 
\cite{Bianchi:2025sjc} contain typos, which have been corrected in the revised arXiv submission. The corrected expressions are displayed here.}
\begin{align}
G_{1;1,1}^{(2)}
={}&
\mathcal{Z}_{\mathrm{FT}}^{(3)}(\epsilon)
\biggl[
\left(N_1^3N_2+N_1N_2^3\right)
\left(
-U_1^{(3)}
-2\,U_2^{(3)}
-8\,U_3^{(3)}
-4\,U_4^{(3)}
+4\,U_5^{(3)}
\right)
\nonumber\\
&+
N_1^2N_2^2
\left(
2\,U_1^{(3)}
+4\,U_2^{(3)}
+16\,U_3^{(3)}
+8\,U_4^{(3)}
-8\,U_5^{(3)}
+4\,U_6^{(3)}
\right)
\nonumber\\
&
-4N_1N_2\,U_6^{(3)}
\biggr]\,.
\label{eq:dimension-one-two-loop-result}
\end{align}
The three lines display separately the color structures $N_1^3N_2+N_1N_2^3$, $N_1^2N_2^2$, and $N_1N_2$. This form makes uniform transcendentality manifest, subject to the assumption above for $U_6^{(3)}$.

Writing $L_n\equiv\operatorname{Li}_n(1/2)$, so that $L_1=\log 2$, the corresponding expansion is
\begin{align}
G_{1;1,1}^{(2)}
={}&-\left(N_1^3N_2+N_1N_2^3-2N_1N_2\right)\zeta_2
\nonumber\\
&+\biggl[
\left(N_1^3N_2+N_1N_2^3\right)
\left(19\zeta_3-26L_1\zeta_2\right)
+N_1^2N_2^2\left(36L_1\zeta_2-55\zeta_3\right)
\nonumber\\
&\qquad+N_1N_2\left(16L_1\zeta_2+17\zeta_3\right)
\biggr]\epsilon
\nonumber\\
&+\biggl[
\left(N_1^3N_2+N_1N_2^3\right)
\left(4L_1^4-98L_1^2\zeta_2+38L_1\zeta_3
+96L_4-\frac{145}{4}\zeta_4\right)
\nonumber\\
&\qquad+N_1^2N_2^2
\left(144L_1^2\zeta_2-110L_1\zeta_3-\frac{463}{2}\zeta_4\right)
\nonumber\\
&\qquad+N_1N_2\left(
-8L_1^4+52L_1^2\zeta_2+34L_1\zeta_3
-192L_4+304\zeta_4\right)
\biggr]\epsilon^2
+O\left(\epsilon^3\right)\,.
\label{eq:dimension-one-epsilon-expansion}
\end{align}

The natural question is whether this structure persists for higher-dimensional operators, whose richer trace structures generate more involved color factors and four-loop propagator integrals.

\subsection{Computational setup}

We generate the relevant Feynman diagrams with \textsc{Qgraf} \cite{Nogueira:1991ex}. Although only a few topological classes occur, the possible chirality assignments of the scalar legs attached to the composite operators lead to substantial combinatorial growth. Relative to the dimension-one case, two structurally new nontrivial contributions appear at two loops.

The integrals are evaluated in momentum space. For operators of dimension $n$, a two-point function can nominally produce an $(n+1)$-loop propagator integral at two-loop perturbative order. Most additional scalar lines, however, remain free and factorize. Consequently, the genuinely interacting momentum integrals required here have at most four loops. We reduce the momentum-space integrals to master integrals using integration-by-parts identities \cite{Tkachov:1981wb,Chetyrkin:1981qh}. Systematic algorithmic approaches to such reductions were developed in \cite{Laporta:2001dd}; here we use the parametric reduction implemented for massless propagators in \textsc{Forcer} \cite{Ruijl:2017cxj}. Their $\epsilon$ expansions are then obtained from the three-dimensional four-loop propagator results \cite{Lee:2015summertime}.

The color algebra is performed in \textsc{Form} \cite{Vermaseren:2000nd,Ruijl:2017dtg,Davies:2026cci}, using an adaptation of the routines in the \texttt{color} package \cite{vanRitbergen:1998pn} to bifundamental $U(N_1)\times U(N_2)$ matter. At higher dimension, the exponential proliferation of scalar contractions becomes the main computational bottleneck. We mitigate it by introducing effective interaction vertices and identifying equivalent contractions before carrying out the color algebra.

Finally, the resulting momentum-space correlators are Fourier transformed back to position space using \eqref{eq:FTconvention} and normalized by the appropriate products of tree-level propagators according to \eqref{eq:perturbative-expansion}.

\subsection{Dimension-two operators}

The planar two-loop two-point function of the single-trace dimension-two operator was previously computed in \cite{Young:2014sia}, as part of the normalization required for an extremal chiral-primary three-point function. We extend that calculation to arbitrary trace structures and exact finite ranks $N_1$ and $N_2$, using the planar single-trace result as a benchmark. 

Dimension-two operators admit new diagrams that connect three scalar lines and produce four-loop momentum integrals.
Additional diagrams involving four scalar lines factorize into two separate two-loop integrals. These diagrams contain only a Chern--Simons gauge-field exchange and vanish by antisymmetry in Feynman gauge. We retained them as a gauge-invariance check but omit them here.

After the integration-by-parts reduction with \textsc{Forcer}, all nonvanishing diagrams reduce to the following master integrals:
\begin{equation}
\begin{aligned}
M_{01}^{(4)}
&\equiv \masterintegralpicN{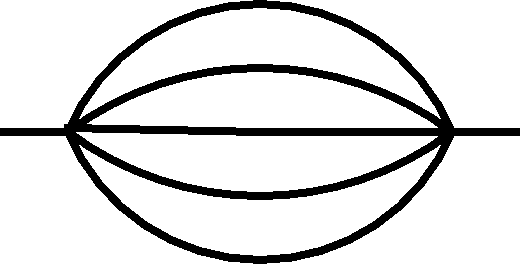}\,,
&\qquad
M_{12}^{(4)}
&\equiv \masterintegralpicN{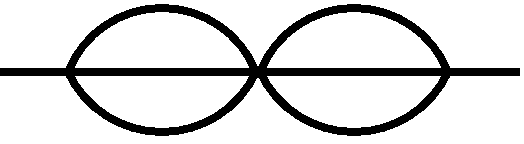}\,,
\\
M_{13}^{(4)}
&\equiv \masterintegralpicN{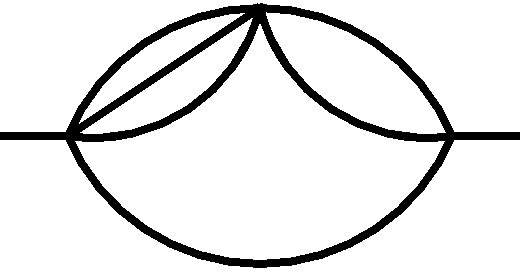}\,,
&\qquad
M_{14}^{(4)}
&\equiv \masterintegralpicN{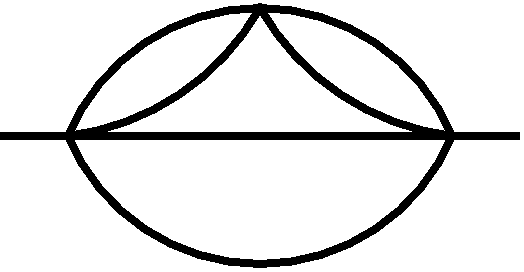}\,,
\\
M_{25}^{(4)}
&\equiv \masterintegralpicN{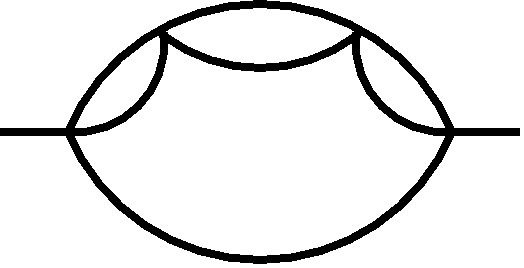}\,,
&\qquad
M_{26}^{(4)}
&\equiv \masterintegralpicN{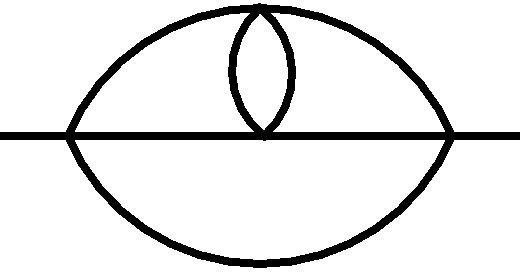}\,,
\\
M_{36}^{(4)}
&\equiv \masterintegralpicN{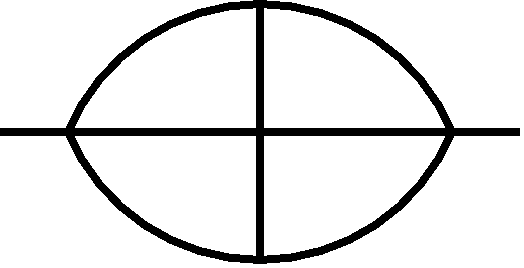}\,,
&\qquad
M_{43}^{(4)}
&\equiv \masterintegralpicN{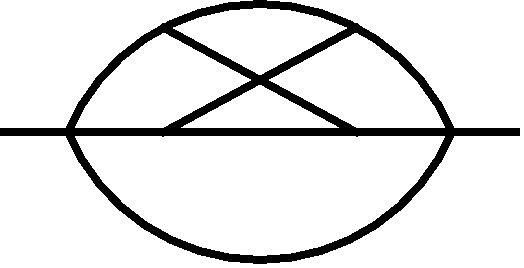}\,.
\end{aligned}
\label{eq:four-loop-master-integrals}
\end{equation}

We follow the nomenclature used in \cite{Baikov:2010hf,Lee:2011jt,Lee:2015summertime}. As representative examples, we first display three color components of the correlator between two single-trace operators. The mixed correlator between a single-trace and a double-trace operator, as well as the correlator between two double-trace operators, is treated analogously. 

Fourier transformation to position space, followed by normalization by the tree-level propagators, yields the overall factor
\begin{align}
\mathcal{Z}'_{\mathrm{FT}}(\epsilon) &\equiv \frac{
\pi^5\,4^{4-4\epsilon}\,e^{-6\gamma_E\epsilon}\,
\Gamma\!\left(\frac32-5\epsilon\right)}
{\Gamma\!\left(\frac12-\epsilon\right)^3\,\Gamma(4\epsilon)}
=
\frac{1-10\epsilon}{\epsilon}\,
\mathcal{Z}_{\mathrm{FT}}(\epsilon)\,,
\\
\mathcal{Z}_{\mathrm{FT}}(\epsilon)
&\equiv
\frac{
\pi^5\,2^{7-8\epsilon}\,e^{-6\gamma_E\epsilon}\,\epsilon\,
\Gamma\!\left(\frac12-5\epsilon\right)}
{\Gamma\!\left(\frac12-\epsilon\right)^3\,\Gamma(4\epsilon)}\,.
\end{align}
Here $\mathcal{Z}_{\mathrm{FT}}(\epsilon)$ is manifestly uniformly transcendental and will be used below.
Using the $\epsilon$ expansions collected in Appendix~\ref{app:expansions}, for the single-trace correlator we find
\begin{align}
\left.G_{2;1,1}^{(2)}\right|_{N_1^4N_2^2}
={}&-4\zeta_2
+\left(76\zeta_3-80L_1\zeta_2\right)\epsilon
\nonumber\\
&+\left(16L_1^4+160\zeta_2L_1^2-304\zeta_3L_1+384L_4-60\zeta_4\right)\epsilon^2
\nonumber\\
&+\biggl(-\frac{384}{5}L_1^5-\frac{256}{3}\zeta_2L_1^3+608\zeta_3L_1^2
-1536L_4L_1-2328\zeta_4L_1
\nonumber\\
&\qquad+1536L_5-\frac{2156}{3}\zeta_2\zeta_3+1950\zeta_5\biggr)\epsilon^3
\nonumber\\
&+\biggl(\frac{2816}{15}L_1^6-\frac{752}{3}\zeta_2L_1^4-\frac{2432}{3}\zeta_3L_1^3
+3072L_4L_1^2+7248\zeta_4L_1^2
\nonumber\\
&\qquad-6144L_5L_1-\frac{7408}{3}\zeta_2\zeta_3L_1-7800\zeta_5L_1
+\frac{7388}{3}\zeta_3^2
\nonumber\\
&\qquad+6144L_6+1152L_4\zeta_2+\frac{3425}{4}\zeta_6-4224\zeta_{-5,-1}\biggr)\epsilon^4
+O\left(\epsilon^5\right)\,,
\end{align}
\begin{align}
\left.G_{2;1,1}^{(2)}\right|_{N_1^3N_2^3}
={}&-8\zeta_2
+\left(224L_1\zeta_2-328\zeta_3\right)\epsilon
\nonumber\\
&+\left(-64L_1^4-448\zeta_2L_1^2+1312\zeta_3L_1-1536L_4-256\zeta_4\right)\epsilon^2
\nonumber\\
&+\biggl(256L_1^5+\frac{256}{3}\zeta_2L_1^3-2624\zeta_3L_1^2
+6144L_4L_1+18304\zeta_4L_1
\nonumber\\
&\qquad+\frac{9896}{3}\zeta_2\zeta_3-24792\zeta_5\biggr)\epsilon^3
\nonumber\\
&+\biggl(-512L_1^6+\frac{4544}{3}\zeta_2L_1^4+\frac{10496}{3}\zeta_3L_1^3
-12288L_4L_1^2
\nonumber\\
&\qquad-79808\zeta_4L_1^2+\frac{52960}{3}\zeta_2\zeta_3L_1+99168\zeta_5L_1
-\frac{116192}{3}\zeta_3^2
\nonumber\\
&\qquad+13824L_4\zeta_2-\frac{158365}{6}\zeta_6+67584\zeta_{-5,-1}\biggr)\epsilon^4
+O\left(\epsilon^5\right)\,,
\end{align}
and
\begin{align}
\left.G_{2;1,1}^{(2)}\right|_{N_1N_2}
={}&8\zeta_2
+\left(16L_1\zeta_2+68\zeta_3\right)\epsilon
\nonumber\\
&+\left(-32L_1^4-32\zeta_2L_1^2-272\zeta_3L_1-768L_4
+1046\zeta_4\right)\epsilon^2
\nonumber\\
&+\biggl(\frac{768}{5}L_1^5-\frac{640}{3}\zeta_2L_1^3
+544\zeta_3L_1^2+3072L_4L_1-2528\zeta_4L_1
\nonumber\\
&\qquad-3072L_5-\frac{1316}{3}\zeta_2\zeta_3+6548\zeta_5\biggr)\epsilon^3
\nonumber\\
&+\biggl(-\frac{5632}{15}L_1^6+\frac{2272}{3}\zeta_2L_1^4
-\frac{2176}{3}\zeta_3L_1^3-6144L_4L_1^2+1792\zeta_4L_1^2
\nonumber\\
&\qquad
+12288L_5L_1-\frac{112}{3}\zeta_2\zeta_3L_1-26192\zeta_5L_1-\frac{6944}{3}\zeta_3^2-12288L_6
\nonumber\\
&
\qquad-5376L_4\zeta_2+\frac{150362}{3}\zeta_6+8448\zeta_{-5,-1}\biggr)\epsilon^4
+O\left(\epsilon^5\right)\,.
\end{align}
Here and below, $\zeta_{a_1,\ldots,a_k}$ denotes a multiple zeta value (MZV) when all indices are positive and an alternating Euler sum when at least one index is negative; for the latter we follow the sign convention of \cite{Bianchi:2025sjc}. In particular, $\zeta_n$ is the ordinary Riemann zeta value.
The $\epsilon\to0$ limit of the planar single-trace correlator
$G_{2;1,1}^{(2)}$ agrees with the result of \cite{Young:2014sia}.
These representative single-trace results, together with the analogous expressions for the mixed single-/double-trace correlator and the correlator between two double-trace operators, exhibit uniform transcendentality through the displayed orders in $\epsilon$.

This property becomes manifest after changing to a uniformly transcendental basis of master integrals. For $M^{(4)}_{01}$, $M^{(4)}_{13}$, $M^{(4)}_{14}$, $M^{(4)}_{25}$, $M^{(4)}_{26}$ and $M^{(4)}_{43}$, the normalization can be inferred by embedding the uniformly transcendental three-loop masters in \eqref{eq:three-loop-ut-first-five} and \eqref{eq:three-loop-ut-six} into four-loop integrals and including the factor associated with the additional momentum integration. The master $M^{(4)}_{12}$ is a product of bubble integrals, so its uniformly transcendental normalization follows directly. We define
\begin{align}
U_{01}^{(4)}
&\equiv
\frac{(1-4\epsilon)(3-10\epsilon)(1-10\epsilon)}{\epsilon^3}
\,M_{01}^{(4)}\,,
\\
U_{12}^{(4)}
&\equiv
\frac{(1-6\epsilon)^2}{\epsilon^2}
\,M_{12}^{(4)}\,,
\\
U_{13}^{(4)}
&\equiv
\frac{(1-6\epsilon)(1-10\epsilon)}{\epsilon^2}
\,M_{13}^{(4)}\,,
\\
U_{14}^{(4)}
&\equiv
\frac{1-10\epsilon}{\epsilon}
\,M_{14}^{(4)}\,,
\\
U_{25}^{(4)}
&\equiv
\frac{1+6\epsilon}{\epsilon}
\,M_{25}^{(4)}\,,
\\
U_{26}^{(4)}
&\equiv
\frac{1+6\epsilon}{\epsilon}
\,M_{26}^{(4)}\,.
\label{eq:four-loop-ut-simple}
\end{align}
The definition of $U_{43}^{(4)}$ instead involves the full combination in \eqref{eq:three-loop-ut-six}:
\begin{align}
U_{43}^{(4)}
\equiv{}&
-\frac{2(1-4\epsilon)(3-10\epsilon)(1-10\epsilon)
\left(3+60\epsilon+172\epsilon^2\right)}
{\epsilon(1+2\epsilon)^3(1+4\epsilon)}
\,M_{01}^{(4)}
\nonumber\\
&+
\frac{14(1-6\epsilon)(1+6\epsilon)(1-10\epsilon)}
{(1+2\epsilon)^2(1+4\epsilon)}
\,M_{13}^{(4)}
\nonumber\\
&+
\frac{192\epsilon(1-10\epsilon)}{(1+2\epsilon)^2}
\,M_{14}^{(4)}
-
\frac{5(1+6\epsilon)^2}{(1+2\epsilon)(1+4\epsilon)}
\,M_{26}^{(4)}
\nonumber\\
&-
\frac{3(1+2\epsilon)^2(1+6\epsilon)(5+6\epsilon)}
{4\epsilon(1+4\epsilon)(1+10\epsilon)(3+10\epsilon)}
\,M_{43}^{(4)}\,.
\label{eq:four-loop-ut-43}
\end{align}
The color components proportional to $N_1^4N_2^2$ and $N_1N_2$ displayed above have manifest uniform transcendentality in this basis:
\begin{align}
\left.G_{2;1,1}^{(2)}\right|_{N_1^4N_2^2}
&= \mathcal{Z}_{\mathrm{FT}}(\epsilon)\left(
-4\,U_{01}^{(4)}-8\,U_{13}^{(4)}-32\,U_{14}^{(4)}
+16\,U_{25}^{(4)}-16\,U_{26}^{(4)}\right)\,,
\\
\left.G_{2;1,1}^{(2)}\right|_{N_1N_2}
&=
\mathcal{Z}_{\mathrm{FT}}(\epsilon)\left(16\,U_{43}^{(4)}\right)\,.
\label{eq:dimension-two-ut-components}
\end{align}

For $M_{36}^{(4)}$, no simple multiplicative normalization enforcing uniform transcendentality is apparent. We instead use the $N_1^3N_2^3$ color component, which contains $M_{36}^{(4)}$ and empirically exhibits uniform transcendentality, to infer the required combination. Its master-integral representation is
\begin{align}
\left.G_{2;1,1}^{(2)}\right|_{N_1^3N_2^3}
={}&
\mathcal{Z}_{\mathrm{FT}}(\epsilon)\biggl[
\frac{8(1+6\epsilon)(1+2\epsilon)^2}
{\epsilon^2(1+10\epsilon)}\,M_{36}^{(4)}
+\frac{8(1+14\epsilon)}{1+6\epsilon}\,U_{01}^{(4)}
\nonumber\\
&\quad-
\frac{8(3+26\epsilon)}{1+10\epsilon}\,U_{12}^{(4)}
+32\,U_{13}^{(4)}
+\frac{512\epsilon}{1+6\epsilon}\,U_{14}^{(4)}\biggr]\,.
\label{eq:dimension-two-M36-component}
\end{align}
Subtracting the $\epsilon$-independent parts of the coefficients whose uniform transcendentality is already manifest leads to the definition
\begin{align}
U_{36}^{(4)}
\equiv{}&
\frac{(1+6\epsilon)(1+2\epsilon)^2}
{\epsilon^2(1+10\epsilon)}\,M_{36}^{(4)}
+\frac{8\epsilon}{1+6\epsilon}\,U_{01}^{(4)}
+
\frac{4\epsilon}{1+10\epsilon}\,U_{12}^{(4)}
+\frac{64\epsilon}{1+6\epsilon}\,U_{14}^{(4)}\,.
\label{eq:four-loop-ut-36}
\end{align}
We extracted the $\epsilon$ expansion of this ansatz numerically with \texttt{SummerTime}, through transcendental weight ten. A PSLQ fit to a minimal basis of uniformly transcendental Euler sums is stable at approximately 2200 digits for all four-loop uniformly transcendental master integrals considered here, providing strong evidence that the combination is uniformly transcendental.

The correlator calculation thus isolates a uniformly transcendental normalization for a topology for which such a choice would otherwise be difficult to identify without embedding the integral in a larger differential-equation system and transforming it to canonical form. Indeed, the coefficients multiplying the top-sector integrals in \eqref{eq:four-loop-ut-36} arise naturally from the corresponding Feynman diagram containing a quartic vertex that couples two scalars to two gauge fields.

The full expansions are supplied in the ancillary data, while the orders required here are displayed in Appendix~\ref{app:expansions}.

We now use this uniformly transcendental basis for higher-dimensional operators.

\subsection{Dimension three and higher}

At dimension three, the same structural diagrams occur as at dimension two, but with richer combinatorics. In particular, the genuinely four-loop diagrams involving three scalar lines may now attach three chiral or three antichiral fields to the same operator. After omitting the factorized integrals that vanish in Feynman gauge, the dimension-three calculation contains all interacting building blocks required at higher dimension; additional scalar lines enter only through free propagator contractions between the two operators.

We exploit this structure by constructing an effective interaction block with six external scalar legs and inserting it into higher-dimensional tree-level contractions. This substantially reduces the complexity of the calculation, although the number of color contractions still grows rapidly with the operator dimension. We compute the complete correlator matrices through dimension four and use them to identify the general pattern. As a separate test, we directly evaluate the complete dimension-five matrix and find full agreement.

All results can be expressed in terms of the three combinations
\begin{align}
\mathcal{A}
&\equiv
\mathcal{Z}_{\mathrm{FT}}(\epsilon)\bigl(
U_{01}^{(4)}
+2\,U_{13}^{(4)}
+8\,U_{14}^{(4)}
-4\,U_{25}^{(4)}
+4\,U_{26}^{(4)}\bigr)\,,
\\
\mathcal{B}
&\equiv
\mathcal{Z}_{\mathrm{FT}}(\epsilon)\bigl(
U_{01}^{(4)}
-3\,U_{12}^{(4)}
+4\,U_{13}^{(4)}
+U_{36}^{(4)}\bigr)\,,
\\
\mathcal{C}
&\equiv
\mathcal{Z}_{\mathrm{FT}}(\epsilon)\bigl(
U_{01}^{(4)}
+2\,U_{13}^{(4)}
+8\,U_{14}^{(4)}
-4\,U_{25}^{(4)}
+4\,U_{26}^{(4)}
-2\,U_{43}^{(4)}\bigr)
\nonumber\\
&=
\mathcal{A}
-2\,\mathcal{Z}_{\mathrm{FT}}(\epsilon)U_{43}^{(4)}\,.
\label{eq:ABC}
\end{align}
At dimension three, the color dependence is most compactly expressed in
terms of the symmetric combinations
\begin{equation}
\begin{aligned}
\sigma&\equiv N_1+N_2\,,
&\qquad
\rho&\equiv N_1N_2\,,
\\
\Delta&\equiv \sigma^2-2\rho-2
=N_1^2+N_2^2-2\,,
&
\Omega&\equiv (\rho+1)^2-\sigma^2
=(N_1^2-1)(N_2^2-1)\,.
\end{aligned}
\label{eq:dimension-three-color-basis}
\end{equation}
The independent components then take the factored form
\begin{align}
G_{3;1,1}^{(2)}
={}&-9\rho\,\Delta
\bigl[\Delta+(\rho+1)(\rho+5)\bigr]\mathcal{A}
\nonumber\\
&+18\rho\,
\bigl[2(\rho+2)\Delta+(\rho+6)\Omega\bigr]\mathcal{B}
-18\rho\,\Omega\mathcal{C}\,,
\nonumber\\[0.5em]
G_{3;1,2}^{(2)}
={}&-18\rho\sigma(\rho+2)\Delta\mathcal{A}
+36\rho\sigma(\rho-1)(\rho+2)\mathcal{B}\,,
\nonumber\\[0.5em]
G_{3;1,3}^{(2)}
={}&-18\rho\,\Delta
\bigl[\Delta+3(\rho+1)\bigr]\mathcal{A}
+36\rho\,
\bigl[(\rho+2)\Delta+2\Omega\bigr]\mathcal{B}
+36\rho\,\Omega\mathcal{C}\,,
\nonumber\\[0.5em]
G_{3;2,2}^{(2)}
={}&-6\rho\,\Delta
\bigl[2\Delta+(\rho+1)(\rho+8)\bigr]\mathcal{A}
\nonumber\\
&+4\rho\,
\bigl[9(\rho+2)\Delta+2(\rho+10)\Omega\bigr]\mathcal{B}
+4\rho(\rho+4)\Omega\mathcal{C}\,,
\nonumber\\[0.5em]
G_{3;2,3}^{(2)}
={}&-18\rho\sigma(\rho+2)\Delta\mathcal{A}
+12\rho\sigma
\bigl[2\Delta+(\rho+4)(\rho-1)\bigr]\mathcal{B}
+24\rho\sigma\Omega\mathcal{C}\,,
\nonumber\\[0.5em]
G_{3;3,3}^{(2)}
={}&-18\rho(\rho+1)(\rho+2)\Delta\mathcal{A}
+36\rho(\rho+2)\Delta\mathcal{B}
+36\rho(\rho+2)\Omega\mathcal{C}\,.
\label{eq:dimension-three-components}
\end{align}
The absence of explicit $\epsilon$-dependent rational coefficients in these combinations makes uniform transcendentality manifest.

At dimension four, two representative diagonal components are
\begin{align}
G_{4;1,1}^{(2)}
={}&-16\rho\,\Delta
\bigl[(5\rho+16)\Delta
+(\rho+1)(\rho^2+19\rho+52)\bigr]\mathcal{A}
\nonumber\\
&+32\rho\,
\bigl[\Delta^2+(7\rho^2+21\rho-19)\Delta
+(\rho^2-1)(\rho^2+24\rho+74)\bigr]\mathcal{B}
\nonumber\\
&-32\rho\,
\bigl[\Delta^2+(2\rho^2+10\rho-3)\Delta
+(\rho^2-1)(5\rho+22)\bigr]\mathcal{C}\,,
\nonumber\\[0.5em]
\dots
\nonumber\\[0.5em]
G_{4;5,5}^{(2)}
={}&-96\rho(\rho+1)(\rho+2)(\rho+3)\Delta\mathcal{A}
+288\rho(\rho+2)(\rho+3)\Delta\mathcal{B}
\nonumber\\
&+96\rho(\rho+2)(\rho+3)
\bigl(2\Omega-\Delta\bigr)\mathcal{C}\,.
\label{eq:dimension-four-components}
\end{align}
The omitted components are also uniformly transcendental and follow from the general formulas presented below.

\section{Planar pattern at arbitrary length}\label{sec:full-planar}

The planar results obtained for operators of lengths $L=4,6,8$ exhibit a simple universal structure when organized in the ordered multi-trace basis of \eqref{eq:ordered-partitions}. At fixed dimension $n$, the $i$th operator is associated with $\mu^{(i)}\,\vdash\,n$ and with the elementary-field trace partition $\lambda^{(i)}=2\mu^{(i)}$. We use the multiplicities $m_r(\mu^{(i)})$ and the centralizer factor $z_{\mu^{(i)}}$ defined in \eqref{eq:zmu}.
In particular, $m_1(\mu^{(i)})$ is the number of unit parts in $\mu^{(i)}$, or equivalently the number of length-two traces in $\lambda^{(i)}$.

In terms of the combinations $\mathcal{A}$, $\mathcal{B}$, and $\mathcal{C}$ defined in \eqref{eq:ABC}, all explicit planar results are reproduced by
\begin{equation}
\begin{aligned}
G^{(2),\mathrm{planar}}_{n;i,i}
={}&
-n\,z_{\mu^{(i)}}(N_1N_2)^{n}
\left(N_1^2+N_2^2\right)\mathcal{A}
\\
&\quad+2z_{\mu^{(i)}}(N_1N_2)^{n+1}
\bigl[
\left(n-m_1(\mu^{(i)})\right)\mathcal{B}
+m_1(\mu^{(i)})\mathcal{C}
\bigr]\,.
\end{aligned}
\label{eq:fullplanar}
\end{equation}
At leading planar order, the correction matrix is diagonal in this basis,
\begin{equation}
G^{(2),\mathrm{planar}}_{n;i,j}=0\,,
\qquad
i\neq j\,,
\label{eq:planar-offdiag}
\end{equation}
for all lengths considered.

The dependence on the trace structure is therefore remarkably constrained. Apart from the overall combinatorial factor $z_{\mu^{(i)}}$, the result only distinguishes length-two traces from longer ones through $m_1(\mu^{(i)})$. In particular, the detailed distribution of the parts $r>1$ does not affect the dynamical combination of master integrals. An analogous simplification was recently found for protected two-point functions in $\mathcal{N}=4$ SYM, where the planar correction is controlled by the number of stress-tensor-multiplet factors in the multi-trace operator \cite{Bianchi:2026tracing}.

As a simple check, at $n=4$ the second partition in \eqref{eq:partition-order-examples} is
\[
\mu^{(2)}=(3,1)\,,
\qquad
\lambda^{(2)}=(6,2)\,.
\]
We find
\[
z_{\mu^{(2)}}=3\,,
\qquad
m_1(\mu^{(2)})=1\,,
\]
so that
\begin{equation}
G^{(2),\mathrm{planar}}_{4;2,2}
=
-12(N_1N_2)^{4}\left(N_1^2+N_2^2\right)\mathcal{A}
+
6(N_1N_2)^{5}\left(3\mathcal{B}+\mathcal{C}\right)\,,
\end{equation}
in agreement with the direct computation.

The factor $z_{\mu^{(i)}}$ is the order of the centralizer of a permutation with cycle type $\mu^{(i)}$, suggesting a natural interpretation of the overall multiplicity in terms of the permutation structure underlying the multi-trace operators. The result \eqref{eq:fullplanar} therefore provides a natural conjecture for the planar correction at arbitrary length.

\section{Finite-rank structure of the two-loop correlators}
\label{sec:finiteN}

The explicit two-loop calculation can be performed without taking the planar limit, thereby retaining the complete dependence on the two gauge-group ranks $N_1$ and $N_2$. The results for operators of lengths $L=4,6,8$ display a remarkably constrained structure, which admits a natural formulation in terms of symmetric-group data.

We use the trace- and Schur-basis conventions of
Section~\ref{sec:tree}. In particular, the exact tree-level metric is given
by \eqref{eq:tree-trace}, with the bifundamental Schur norms defined in
\eqref{eq:FRdef}.

The same three master-integral combinations $\mathcal{A}$, $\mathcal{B}$, and $\mathcal{C}$ defined in \eqref{eq:ABC} are sufficient to describe the complete finite-rank result. No additional dynamical structures appear when non-planar color contractions are included.

For all the cases explicitly computed, the full two-loop correction can be organized as
\begin{equation}
\begin{aligned}
G^{(2)}_{n;i,j}
={}&
\left[
-n\left(N_1^2+N_2^2-2\right)\mathcal{A}
+n\left(N_1N_2-1\right)
\left(\mathcal{B}+\mathcal{C}\right)
\right]
G^{(0)}_{n;i,j}
+\frac{\mathcal{B}-\mathcal{C}}{2}
K^{(n)}_{i,j}\,.
\end{aligned}
\label{eq:finiteN-master}
\end{equation}
Thus, two independent combinations are entirely proportional to the exact tree-level metric, while all genuinely new finite-rank information is contained in a single kernel $K^{(n)}$.

The structure of this kernel becomes particularly transparent after transforming to the Schur basis, where the tree-level metric is diagonal as in \eqref{eq:tree-schur}. We further define the total content of a Young diagram,
\begin{equation}
c_R
=
\sum_{(a,b)\in R}(b-a)\,.
\label{eq:content}
\end{equation}
The finite-rank data for $n=2,3,4$ are reproduced by
\begin{equation}
K^{(n)}_{R,S}
=
\delta_{RS}F_R
\left[
2n\left(N_1N_2+1\right)
+4\left(N_1+N_2\right)c_R
\right]
-
4F_R F_S
\sum_{\substack{T\,\vdash\,n-1\\
T\nearrow R,\;T\nearrow S}}
\frac{1}{F_T}\,.
\label{eq:Kschur}
\end{equation}
Here $F_R\equiv F_R(N_1,N_2)$, and $T\nearrow R$ denotes the addition of one box to the Young diagram $T$ to obtain $R$.
Although \eqref{eq:Kschur} and its matrix form \eqref{eq:Kmatrix} involve $F_T^{-1}$, each ratio $F_RF_S/F_T$ is polynomial in the ranks. These formulas are therefore understood at generic $N_1,N_2$, with their values at ranks for which some Schur norms vanish obtained by polynomial continuation.

The expression \eqref{eq:Kschur} shows that the non-diagonal part of the two-loop correction has a simple geometrical interpretation on the Young lattice. For $R\neq S$, the matrix element $K^{(n)}_{R,S}$ vanishes unless the two Young diagrams share a common predecessor $T\,\vdash\,n-1$. Equivalently, $R$ and $S$ must differ by the displacement of a single box. Whenever such a common diagram exists,
\begin{equation}
K^{(n)}_{R,S}
=
-4\,\frac{F_R F_S}{F_T}\,.
\label{eq:Koffdiag}
\end{equation}
The finite-rank mixing is therefore local on the Young graph. A closely related one-box mixing rule appears in the finite-rank one-loop problem in $\mathcal{N}=4$ SYM \cite{Brown:2008permutations}, while broader Young-diagram lattice structures arise in restricted-Schur descriptions of non-planar anomalous dimensions \cite{DeComarmond:2010spectra}.

A compact matrix representation of the same result can be obtained by introducing the diagonal matrices
\begin{equation}
(D_n)_{RR}=F_R\,,
\qquad
(C_n)_{RR}=c_R\,,
\end{equation}
together with the incidence matrix
\begin{equation}
(U_n)_{TR}
=
\begin{cases}
1\,, & T\nearrow R\,,\\
0\,, & \text{otherwise}\,,
\end{cases}
\qquad
T\,\vdash\,n-1,\quad R\,\vdash\,n\,.
\end{equation}
The kernel then reads
\begin{equation}
K_n
=
2n\left(N_1N_2+1\right)D_n
+
4\left(N_1+N_2\right)C_n D_n
-
4D_n U_n^{\mathrm T}D_{n-1}^{-1}U_n D_n\,.
\label{eq:Kmatrix}
\end{equation}
The result in the original multi-trace basis follows by the character transformation
\begin{equation}
K^{(n)}_{i,j}
=
\sum_{R,S\,\vdash\,n}
\chi_R\!\left(\mu^{(i)}\right)
\chi_S\!\left(\mu^{(j)}\right)
K^{(n)}_{R,S}\,.
\label{eq:Ktrace}
\end{equation}

The decomposition~\eqref{eq:finiteN-master} makes the separation between dynamics and finite-rank combinatorics manifest. All dependence on the loop integrations is contained in the three universal quantities $\mathcal{A}$, $\mathcal{B}$, and $\mathcal{C}$. The terms proportional to $\mathcal{A}$ and $\mathcal{B}+\mathcal{C}$ are controlled entirely by the exact free two-point function, whereas the combination $\mathcal{B}-\mathcal{C}$ probes a new operator on the space of Young diagrams, whose off-diagonal action connects \emph{nearest neighbours} on the Young lattice, namely diagrams related by a single-box move.

It is useful to verify that the finite-rank expression consistently reproduces the planar result discussed above. Using the centralizer factor $z_{\mu^{(i)}}$, the multiplicity $m_1(\mu^{(i)})$, and the tree-level limit \eqref{eq:tree-planar}, the only additional ingredient is the leading behavior of the kernel,
\begin{equation}
K^{(n)}_{i,j}
=
2z_{\mu^{(i)}}
\left(n-2m_1(\mu^{(i)})\right)
N_1N_2
\left(N_1N_2\right)^n
\delta_{ij}
+\text{subleading in }N_1,N_2\,.
\label{eq:K-planar-limit}
\end{equation}
Substitution into \eqref{eq:finiteN-master} gives the leading planar contribution
\begin{equation}
\begin{aligned}
G^{(2),\mathrm{planar}}_{n;i,i}
={}&
z_{\mu^{(i)}}\left(N_1N_2\right)^n
\Bigl[
-n\left(N_1^2+N_2^2\right)\mathcal{A}
\\
&\hphantom{z_{\mu^{(i)}}\left(N_1N_2\right)^n\Bigl[}
+2N_1N_2
\left(
\left(n-m_1(\mu^{(i)})\right)\mathcal{B}
+m_1(\mu^{(i)})\mathcal{C}
\right)
\Bigr]\,,
\end{aligned}
\label{eq:finiteN-to-planar}
\end{equation}
in exact agreement with the planar pattern obtained above.

The expressions \eqref{eq:finiteN-master} and \eqref{eq:Kschur} were inferred
from the complete finite-$(N_1,N_2)$ results through dimension four. As an
independent check, we computed the full dimension-five correlator matrix
directly and found exact agreement with their prediction. The ancillary \textsc{Mathematica}
notebook \texttt{examples.nb} contains the corresponding tree-level and
two-loop matrices, together with an explicit implementation of the formulas
presented in this section.

\subsection{Final result and all-length prediction}
\label{sec:integrated-finiteN}

The preceding formulas separate the universal color structure from the dynamical quantities $\mathcal{A}$, $\mathcal{B}$, and $\mathcal{C}$. Their three-dimensional limits are
\begin{equation}
\mathcal{A}\xrightarrow[\epsilon\to0]{}\zeta_2\,,
\qquad
\mathcal{B}\xrightarrow[\epsilon\to0]{}-\zeta_2\,,
\qquad
\mathcal{C}\xrightarrow[\epsilon\to0]{}0.
\label{eq:ABC-integrated}
\end{equation}

Substitution into \eqref{eq:finiteN-master} gives the central prediction of this work. For arbitrary operator dimension $n$, finite ranks $(N_1,N_2)$, and any pair of trace structures $\mu^{(i)},\mu^{(j)}\,\vdash\,n$, the proposed physical two-loop correction is
\begin{equation}
\boxed{
G^{(2)}_{n;i,j}
=
-n\,\zeta_2
\left(
N_1^2+N_1N_2+N_2^2-3
\right)
G^{(0)}_{n;i,j}
-\frac{\zeta_2}{2}\,K^{(n)}_{i,j}
}
\label{eq:integrated-trace-result}
\end{equation}
where the exact tree-level metric and the universal kernel are given in \eqref{eq:tree-trace} and \eqref{eq:Ktrace}, respectively.

Equivalently, using the Schur norm $F_R$ and total content $c_R$ defined in \eqref{eq:FRdef} and \eqref{eq:content}, the prediction reads
\begin{equation}
\boxed{\begin{aligned}
G^{(2)}_{n;R,S}
={}&
-\zeta_2\,
\delta_{RS}\,F_R
\left[
n\left((N_1+N_2)^2-2\right)
+
2(N_1+N_2)c_R
\right]
\\
&+
2\zeta_2\,
F_R F_S
\sum_{\substack{
T\,\vdash\,n-1\\
T\nearrow R,\;T\nearrow S
}}
\frac{1}{F_T}
\end{aligned}}
\label{eq:integrated-schur-result}
\end{equation}
This form displays the prediction particularly clearly: the diagonal term is fixed by the Schur norm and the total content of the diagram, while the off-diagonal term connects only diagrams with a common predecessor on the Young lattice. The planar limit follows directly from \eqref{eq:tree-planar} and \eqref{eq:K-planar-limit} and reproduces \eqref{eq:fullplanar} after inserting \eqref{eq:ABC-integrated}. Thus, the boxed formulas provide the complete predicted value of the two-loop correction at arbitrary operator length and trace structure, with no further dynamical input.

The formula also applies smoothly at $n=1$, although the corresponding
master-integral representation is special because it involves only
three-loop propagator integrals. For $n=1$ and $n=2$, it reproduces the
$\epsilon\to0$ limits of the explicit results in
\eqref{eq:dimension-one-epsilon-expansion},
\eqref{eq:dimension-two-ut-components}, and
\eqref{eq:dimension-two-M36-component}. As a higher-dimensional
illustration, for $n=4$ the single-trace entry is
\begin{align}
G_{4;1,1}^{(2)}
={}&-\frac{8}{3}\,N_1N_2\biggl[
\left(N_1^5N_2^3+N_1^3N_2^5\right)
+2N_1^4N_2^4
\nonumber\\
&\hspace{5em}{}
+5\left(N_1^5N_2+N_1N_2^5\right)
+34\left(N_1^4N_2^2+N_1^2N_2^4\right)
+56N_1^3N_2^3
\nonumber\\
&\hspace{5em}{}
+18\left(N_1^4+N_2^4\right)
+93\left(N_1^3N_2+N_1N_2^3\right)
+114N_1^2N_2^2
\nonumber\\
&\hspace{5em}{}
-58\left(N_1^2+N_2^2\right)
-254N_1N_2-104
\biggr]\pi^2\,.
\end{align}
As a more intricate finite-rank example, consider $n=7$ and the
diagonal entry associated with the tenth partition in the ordering
\eqref{eq:ordered-partitions}, $\mu^{(10)}=(3,2,1,1)$.
To make the symmetry under $N_1\leftrightarrow N_2$ manifest, define
\begin{equation}
S_m\equiv N_1^m+N_2^m\,,
\qquad
\rho\equiv N_1N_2\,.
\end{equation}
The result then decomposes into symmetric color components as
\begin{equation}
\begin{aligned}
G^{(2)}_{7;10,10}
={}&-2\rho\pi^2\biggl[
756S_8
\\
&\quad
+\left(210240+51408\rho+1883\rho^2\right)S_6
\\
&\quad
+\left(
2493540+2906276\rho+559714\rho^2
+26540\rho^3+294\rho^4
\right)S_4
\\
&\quad
+\left(
-9605400+5121172\rho+9119817\rho^2
+1778698\rho^3
\right.
\\
&\hspace{7em}\left.
{}+92353\rho^4+1422\rho^5+7\rho^6
\right)S_2
\\
&\quad
-7916736-31681176\rho-264174\rho^2
+11780810\rho^3
\\
&\hspace{7em}
{}+2432456\rho^4+129324\rho^5
+2046\rho^6+10\rho^7
\biggr]\,.
\end{aligned}
\label{eq:n-seven-partition-ten-example}
\end{equation}

For the maximal multi-trace component the formula collapses to
\begin{equation}
\begin{aligned}
G^{(2)}_{n;p(n),p(n)}
={}&
-n\,\zeta_2\,
\frac{N_1N_2+n}{N_1N_2+1}\,
\left(N_1^2+N_2^2-2\right)\,
G^{(0)}_{n;p(n),p(n)}
\\
={}&
-n\,n!\,\zeta_2\,N_1N_2\,
\left(N_1^2+N_2^2-2\right)\,
\prod_{r=2}^{n}
\left(N_1N_2+r\right)\,.
\end{aligned}
\end{equation}

\section{Conclusions}

We have studied two-loop two-point functions of protected scalar operators in ABJM theory, retaining arbitrary multi-trace structures and the exact dependence on the ranks $N_1$ and $N_2$. The calculation combines diagram generation, four-loop propagator reduction, and finite-rank color algebra. The all-length structures emerge from the complete results through dimension four and pass a nontrivial independent test against the directly computed full matrix at dimension five.

A central outcome is the construction of a uniformly transcendental basis for the four-loop master integrals required by these correlators. Most normalizations follow from lower-loop uniformly transcendental integrals or factorized bubble representations. The remaining topology is fixed by demanding uniform transcendentality of a protected correlator, yielding the combination $U_{36}^{(4)}$ in \eqref{eq:four-loop-ut-36}. Its high-precision expansion provides a nontrivial consistency check of this procedure.

The complete higher-dimensional results depend on only three universal dynamical quantities, $\mathcal{A}$, $\mathcal{B}$, and $\mathcal{C}$. In the planar limit, the trace-structure dependence of the $i$th operator is encoded by the centralizer factor $z_{\mu^{(i)}}$ and the number $m_1(\mu^{(i)})$ of length-two traces. At finite rank, the correction separates into a part proportional to the exact tree-level metric and a single additional kernel. In the Schur basis, the off-diagonal action of this kernel connects Young diagrams that share a common predecessor, making the mixing local on the Young lattice.

The planar simplification closely parallels the structure recently found for higher-dimensional protected correlators in $\mathcal{N}=4$ SYM, where the dependence on the multi-trace operator is likewise controlled by the number of stress-tensor-multiplet factors \cite{Bianchi:2026tracing}. There is, however, an important difference in how uniform transcendentality is realized. At the first nontrivial perturbative order considered here, the ABJM correlators are uniformly transcendental without any subtraction at the level of the observable, whereas the higher-dimensional $\mathcal{N}=4$ SYM two-point functions require suitable subtractions to preserve uniform transcendentality. Since the two-loop ABJM contribution is the first nonvanishing correction and, in this respect, plays a role analogous to the one-loop correction in $\mathcal{N}=4$ SYM, it is natural to ask whether analogous subtractions will become necessary at higher loop order in ABJM\@. Beyond these similarities and differences, the finite-rank ABJM result raises a further question: whether the nearest-neighbour kernel on the Young lattice admits a more intrinsic algebraic or dynamical interpretation. It would also be interesting to determine whether a comparable finite-rank organization governs protected correlators in $\mathcal{N}=4$ SYM at two loops.

A complementary direction would be to derive the physical three-dimensional correlator matrices directly from supersymmetric localization. Higgs-branch operators in three-dimensional theories with $\mathcal{N}\geq4$ supersymmetry admit a one-dimensional topological description, and derivatives of the mass-deformed $S^3$ partition function generate their integrated correlators; the relevant cohomological relation applies in particular to ABJM theory \cite{Kapustin:2009kz,Dedushenko:2016jxl,Gorini:2020new,Guerrini:2021zuk}. This matrix-model technology has already reproduced the finite-rank two-loop dimension-one ABJM correlator, while higher mass derivatives have been used to extract data associated with dimension-two topological operators \cite{Gorini:2020new,Armanini:2024loc}. It should therefore be possible in principle to recover the $\epsilon\to0$ limit of the correlator matrices studied here. An instructive comparison is provided by the localization analysis of protected two-point functions in $\mathcal{N}=4$ SYM in $d=4-2\epsilon$ \cite{Georgoudis:2026loc}, which requires an analytic continuation of localization to the non-integer-dimensional sphere $S^{4-2\epsilon}$. In the present ABJM case, by contrast, the two-loop correlators have a nontrivial finite limit as $\epsilon\to0$, so their physical normalization can be tested by localization directly on $S^3$, without continuing the dimension of the sphere. The main additional challenge is to introduce sufficiently many independent supersymmetric sources, or equivalently a sufficiently resolved basis of composite insertions in the one-dimensional theory, to disentangle operators of equal dimension with different trace structures. Such a construction would provide a nonperturbative test of the finite-rank formulas, although the regulator-dependent higher-order $\epsilon$ expansion, and hence the full uniform-transcendentality information discussed here, is not directly accessible through standard three-sphere localization.

These observations suggest that protected ABJM correlators admit a compact organization at arbitrary length, despite the rapid growth of their trace and color combinatorics. It would be useful to derive the planar and finite-rank formulas directly from an effective operator acting on the half-BPS sector, to test them at higher lengths, and to investigate whether the correlator-based construction of uniformly transcendental master integrals extends to higher perturbative orders.

\acknowledgments

This work was supported by Fondo Nacional de Desarrollo Cient\'{\i}fico y Tecnol\'ogico, through Fondecyt Exploraci\'on 13250014.

\vfill
\newpage

\appendix

\section{Master integral expansions}\label{app:expansions}

Momentum integrals are defined according to the measure
\begin{equation}
\int_k
\equiv
\left(\frac{e^{\gamma_E}}{4\pi}\right)^{\!\epsilon}
\int\frac{d^d k}{(2\pi)^d}
=
\frac{1}{(4\pi)^{3/2}}
\int\frac{e^{\gamma_E\epsilon}\,d^d k}{\pi^{d/2}}\,,
\qquad d=3-2\epsilon\,.
\label{eq:loop-measure}
\end{equation}
The factor $e^{\gamma_E\epsilon}$ removes Euler--Mascheroni constants from the expansions, while the normalization by $\pi^{d/2}$ removes logarithms of $\pi$.
To avoid spurious Euler--Mascheroni constants and logarithms of $\pi$ in intermediate position-space expressions, we use the Fourier-transform convention
\begin{equation}
\pi^{\frac{d-3}{2}}\,
e^{\left(2\alpha+\frac{d-3}{2}\right)\gamma_E}
\int\frac{d^d p}{(2\pi)^d}\,
e^{i p\cdot x}\,(p^2)^\alpha
=
\frac{4^\alpha
e^{\left(2\alpha+\frac{d-3}{2}\right)\gamma_E}}
{\pi^{3/2}}
\frac{\Gamma\!\left(\frac d2+\alpha\right)}
{\Gamma(-\alpha)}
\frac{1}{(x^2)^{\frac d2+\alpha}}\,.
\label{eq:FTconvention}
\end{equation}
We use
$L_n\equiv\operatorname{Li}_n(1/2)$, in particular $L_1=\log 2$.
As in the main text, $\zeta_{a_1,\ldots,a_k}$ denotes a multiple zeta value (MZV) when all indices are positive and an alternating Euler sum when at least one index is negative; for the latter we follow the sign convention of \cite{Bianchi:2025sjc}. In particular, $\zeta_n$ denotes the ordinary Riemann zeta value.
The complete Laurent expansions of the four-loop masters through transcendental weight ten are supplied in the ancillary file \texttt{expansionsUT4L.m}. They are based partly on \cite{Bianchi:2025sjc} and include the additional expansion of $U^{(4)}_{36}$ obtained and used to test uniform transcendentality in this work. To keep the expressions compact and make the normalization explicit, we define the rescaled masters $\widetilde{U}\equiv(8\pi)^4U$. Below we display a reduced snapshot of their expansions, evaluated at $p^2=1$ and truncated at $O(\epsilon^4)$. The masters normalized according to \eqref{eq:loop-measure} are recovered as $U=(8\pi)^{-4}\widetilde{U}$.
\begin{alignat}{2}
&\widetilde{U}_{01}^{(4)} &&= -\frac{1}{\epsilon ^4} +\frac{22 \zeta _2}{\epsilon ^2} +\frac{904 \zeta _3}{3 \epsilon } +1656 \zeta
   _4 \notag\\
& &&\quad+\bigl(\frac{97744 \zeta _5}{5}-\frac{19888 \zeta _2 \zeta _3}{3}\bigr) \epsilon +\bigl(84043 \zeta
   _6-\frac{408608 \zeta _3^2}{9}\bigr) \epsilon ^2 \notag\\
& &&\quad+\bigl(-499008 \zeta _3 \zeta _4-\frac{2150368 \zeta _2 \zeta _5}{5} +\frac{9937624 \zeta _7}{7}\bigr)
   \epsilon ^3 +O\bigl(\epsilon^4\bigr)\,, \\[0.5em]
&\widetilde{U}_{12}^{(4)} &&= \frac{1}{\epsilon ^4} -\frac{14 \zeta _2}{\epsilon ^2} -\frac{352 \zeta _3}{3 \epsilon } -332 \zeta _4
   \notag\\
& &&\quad+\bigl(\frac{4928 \zeta _2 \zeta _3}{3}-\frac{14944 \zeta _5}{5}\bigr) \epsilon +\bigl(\frac{61952 \zeta
   _3^2}{9}-3089 \zeta _6\bigr) \epsilon ^2 \notag\\
& &&\quad+\bigl(\frac{116864 \zeta _3 \zeta _4}{3}+\frac{209216 \zeta _2 \zeta _5}{5} -\frac{554992 \zeta
   _7}{7}\bigr) \epsilon ^3 +O\bigl(\epsilon^4\bigr)\,, \\[0.5em]
&\widetilde{U}_{13}^{(4)} &&= \frac{1}{2 \epsilon ^4} -\frac{19 \zeta _2}{\epsilon ^2} -\frac{428 \zeta _3}{3 \epsilon } -496 \zeta
   _4 \notag\\
& &&\quad+\Bigl(\frac{16264 \zeta _2 \zeta _3}{3}-\frac{48392 \zeta _5}{5}\Bigr) \epsilon +\Bigl(\frac{183184 \zeta
   _3^2}{9}-\frac{31943 \zeta _6}{2}\Bigr) \epsilon ^2 \notag\\
& &&\quad+\Bigl(\frac{424576 \zeta _3 \zeta _4}{3}+\frac{1838896 \zeta _2 \zeta _5}{5} -\frac{4960748 \zeta
   _7}{7}\Bigr) \epsilon ^3 +O\bigl(\epsilon^4\bigr)\,, \\[0.5em]
&\widetilde{U}_{14}^{(4)} &&= \frac{3 \zeta _2}{\epsilon ^2} +\frac{12 L_1 \zeta _2}{\epsilon } \notag\\
& &&\quad+\bigl(24 L_1^2 \zeta _2 -75 \zeta _4\bigr) +\bigl(32 \zeta _2 L_1^3-120 \zeta _2^2 L_1 -730 \zeta _2
   \zeta _3\bigr) \epsilon \notag\\
& &&\quad+\bigl(32 \zeta _2 L_1^4-240 \zeta _2^2 L_1^2-2920 \zeta _2 \zeta _3 L_1-5499 \zeta _2 \zeta _4\bigr)
   \epsilon ^2 \notag\\
& &&\quad+\bigl(\frac{128}{5} \zeta _2 L_1^5 -320 \zeta _2^2 L_1^3 -5840 \zeta _2 \zeta _3 L_1^2 -21996 \zeta _2
   \zeta _4 L_1 \notag\\
& &&\qquad +7300 \zeta _2^2 \zeta _3 -\frac{270822 \zeta _2 \zeta _5}{5}\bigr) \epsilon ^3 +O\bigl(\epsilon^4\bigr)\,, \\[0.5em]
&\widetilde{U}_{25}^{(4)} &&= -\frac{3 \zeta _2}{\epsilon ^2} -\frac{12 \bigl(L_1 \zeta _2\bigr)}{\epsilon } \notag\\
& &&\quad+\bigl(195 \zeta _4-24 L_1^2 \zeta _2\bigr) +\bigl(-32 \zeta _2 L_1^3 +312 \zeta _2^2 L_1+682 \zeta _2
   \zeta _3\bigr) \epsilon \notag\\
& &&\quad+\bigl(-32 \zeta _2 L_1^4+624 \zeta _2^2 L_1^2+2728 \zeta _2 \zeta _3 L_1 +4947 \zeta _2 \zeta _4\bigr)
   \epsilon ^2 \notag\\
& &&\quad+\bigl(-\frac{128}{5} \zeta _2 L_1^5 +832 \zeta _2^2 L_1^3 +5456 \zeta _2 \zeta _3 L_1^2 +19788 \zeta _2
   \zeta _4 L_1 \notag\\
& &&\qquad -17732 \zeta _2^2 \zeta _3 +\frac{267942 \zeta _2 \zeta _5}{5}\bigr) \epsilon ^3 +O\bigl(\epsilon^4\bigr)\,, \\[0.5em]
&\widetilde{U}_{26}^{(4)} &&= -\frac{3 \zeta _2}{\epsilon ^2} +\frac{12 L_1 \zeta _2-42 \zeta _3}{\epsilon } \notag\\
& &&\quad+\bigl(-8 L_1^4+24 \zeta _2 L_1^2-192 L_4+84 \zeta _4\bigr) \notag\\
& &&\quad+\bigl(\frac{32 L_1^5}{5} -32 \zeta _2 L_1^3 -336 \zeta _4 L_1 -768 L_5 +1786 \zeta _2 \zeta _3 -1023
   \zeta _5\bigr) \epsilon \notag\\
& &&\quad+\bigl(-\frac{64 L_1^6}{15} +208 \zeta _2 L_1^4 -1968 \zeta _4 L_1^2 -3448 \zeta _2 \zeta _3 L_1 +11570
   \zeta _3^2 \notag\\
& &&\qquad -3072 L_6 +4224 L_4 \zeta _2 +\frac{31731 \zeta _6}{4} +2112 \zeta _{-5,-1}\bigr) \epsilon ^2 \notag\\
& &&\quad+\bigl(\frac{256 L_1^7}{105} -\frac{832}{5} \zeta _2 L_1^5 +\frac{7232}{3} \zeta _3 L_1^4 +2624 \zeta _4
   L_1^3 -7568 \zeta _2 \zeta _3 L_1^2 \notag\\
& &&\qquad +8184 \zeta _5 L_1^2 +\frac{38400}{7} \zeta _3^2 L_1 -\frac{271401 \zeta _6 L_1}{7} -\frac{30720}{7}
   \zeta _{-5,-1} L_1 \notag\\
& &&\qquad -12288 L_7 +16896 L_5 \zeta _2 +57856 L_4 \zeta _3 +\frac{215672 \zeta _3 \zeta _4}{7} \notag\\
& &&\qquad +\frac{320772 \zeta _2 \zeta _5}{5} -6000 \zeta _7 +\frac{30720}{7} \zeta _{-5,1,1} -\frac{28416}{7}
   \zeta _{5,-1,-1}\bigr) \epsilon ^3 +O\bigl(\epsilon^4\bigr)\,, \\[0.5em]
&\widetilde{U}_{36}^{(4)} &&= \frac{2}{\epsilon^4}
+\frac{4\zeta_2}{\epsilon^2}
+\frac{192L_1\zeta_2-\frac{1232}{3}\zeta_3}{\epsilon}
\notag\\
& &&\quad+\bigl(
-64L_1^4+384\zeta_2L_1^2-1536L_4-424\zeta_4
\bigr)
\notag\\
& &&\quad+\bigl(
\frac{11456}{3}\zeta_2\zeta_3+5280L_1\zeta_4
-\frac{72968}{5}\zeta_5
\bigr)\epsilon
\notag\\
& &&\quad+\bigl(
2176\zeta_2L_1^4-32640\zeta_4L_1^2
-27904\zeta_2\zeta_3L_1+\frac{416464}{9}\zeta_3^2
\notag\\
& &&\qquad
+52224L_4\zeta_2-\frac{71390}{3}\zeta_6
+67584\zeta_{-5,-1}
\bigr)\epsilon^2
\notag\\
& &&\quad+\bigl(
\frac{12288}{5}\zeta_2L_1^5+\frac{33280}{3}\zeta_3L_1^4
-30720\zeta_4L_1^3-34304\zeta_2\zeta_3L_1^2
\notag\\
& &&\qquad
-30720\zeta_3^2L_1+73728L_4\zeta_2L_1
+188208\zeta_6L_1+24576\zeta_{-5,-1}L_1
\notag\\
& &&\qquad
+73728L_5\zeta_2+266240L_4\zeta_3
+\frac{1377520}{3}\zeta_3\zeta_4
-\frac{1104016}{5}\zeta_2\zeta_5
\notag\\
& &&\qquad
-\frac{1312844}{7}\zeta_7
-24576\zeta_{-5,1,1}
-24576\zeta_{5,-1,-1}
\bigr)\epsilon^3
+O\bigl(\epsilon^4\bigr)\,, \\[0.5em]
&\widetilde{U}_{43}^{(4)} &&= \frac{4 \zeta _2}{\epsilon ^2} +\frac{24 L_1 \zeta _2+34 \zeta _3}{\epsilon } \notag\\
& &&\quad+\bigl(-16 L_1^4+48 \zeta _2 L_1^2-384 L_4+273 \zeta _4\bigr) \notag\\
& &&\quad+\bigl(\frac{64 L_1^5}{5} -64 \zeta _2 L_1^3 -672 \zeta _4 L_1 -1536 L_5 -\frac{6880 \zeta _2 \zeta _3}{3}
   \notag\\
& &&\qquad +3274 \zeta _5\bigr) \epsilon \notag\\
& &&\quad+\bigl(-\frac{128 L_1^6}{15} +352 \zeta _2 L_1^4 -2976 \zeta _4 L_1^2 -8240 \zeta _2 \zeta _3 L_1
   -\frac{34684 \zeta _3^2}{3} \notag\\
& &&\qquad -6144 L_6 +6912 L_4 \zeta _2 -\frac{24695 \zeta _6}{3} +4224 \zeta _{-5,-1}\bigr) \epsilon ^2 \notag\\
& &&\quad+\bigl(\frac{512 L_1^7}{105} -\frac{4736}{5} \zeta _2 L_1^5 +\frac{17152}{3} \zeta _3 L_1^4 +12928 \zeta
   _4 L_1^3 \notag\\
& &&\qquad -28576 \zeta _2 \zeta _3 L_1^2 +16368 \zeta _5 L_1^2 +\frac{184320}{7} \zeta _3^2 L_1 -18432 L_4 \zeta
   _2 L_1 \notag\\
& &&\qquad -\frac{393618 \zeta _6 L_1}{7} -\frac{147456}{7} \zeta _{-5,-1} L_1 -24576 L_7 +15360 L_5 \zeta _2
   \notag\\
& &&\qquad +137216 L_4 \zeta _3 -\frac{1171936 \zeta _3 \zeta _4}{7} -\frac{726496 \zeta _2 \zeta _5}{5} +115816
   \zeta _7 \notag\\
& &&\qquad +\frac{147456}{7} \zeta _{-5,1,1} +\frac{29184}{7} \zeta _{5,-1,-1}\bigr) \epsilon ^3 +O\bigl(\epsilon^4\bigr)\,.
\end{alignat}

\bibliographystyle{JHEP}
\bibliography{biblio3}

@article{Beisert:2010jr,
	Archiveprefix = {arXiv},
	Author = {Beisert, Niklas and others},
	Doi = {10.1007/s11005-011-0529-2},
	Eprint = {1012.3982},
	Journal = {Lett. Math. Phys.},
	Pages = {3-32},
	Primaryclass = {hep-th},
	Reportnumber = {AEI-2010-175, CERN-PH-TH-2010-306, HU-EP-10-87, HU-MATH-2010-22, KCL-MTH-10-10, UMTG-270, UUITP-41-10},
	Slaccitation = {%%CITATION = ARXIV:1012.3982;%%},
	Title = {{Review of AdS/CFT Integrability: An Overview}},
	Volume = {99},
	Year = {2012}}

@article{Young:2014lka,
      author         = "Young, Donovan",
      title          = "{ABJ(M) Chiral Primary Three-Point Function at
                        Two-loops}",
      journal        = "JHEP",
      volume         = "07",
      year           = "2014",
      pages          = "120",
      doi            = "10.1007/JHEP07(2014)120",
      eprint         = "1404.1117",
      archivePrefix  = "arXiv",
      primaryClass   = "hep-th",
      reportNumber   = "QMUL-PH-14-10",
      SLACcitation   = "%%CITATION = ARXIV:1404.1117;%%"
}

@article{Young:2013formfactors,
      author         = "Young, Donovan",
      title          = "{Form Factors of Chiral Primary Operators at Two Loops in ABJ(M)}",
      journal        = "JHEP",
      volume         = "06",
      year           = "2013",
      pages          = "049",
      doi            = "10.1007/JHEP06(2013)049",
      eprint         = "1305.2422",
      archivePrefix  = "arXiv",
      primaryClass   = "hep-th"
}

@article{Brandhuber:2013sudakov,
      author         = "Brandhuber, Andreas and Gurdogan, Omer and Korres, Dimitrios and Mooney, Robert and Travaglini, Gabriele",
      title          = "{Two-loop Sudakov Form Factor in ABJM}",
      journal        = "JHEP",
      volume         = "11",
      year           = "2013",
      pages          = "022",
      doi            = "10.1007/JHEP11(2013)022",
      eprint         = "1305.2421",
      archivePrefix  = "arXiv",
      primaryClass   = "hep-th"
}

@article{Bianchi:2013lightlike,
      author         = "Bianchi, Marco S. and Giribet, Gaston and Leoni, Matias and Penati, Silvia",
      title          = "{Light-like Wilson loops in ABJM and maximal transcendentality}",
      journal        = "JHEP",
      volume         = "08",
      year           = "2013",
      pages          = "111",
      doi            = "10.1007/JHEP08(2013)111",
      eprint         = "1304.6085",
      archivePrefix  = "arXiv",
      primaryClass   = "hep-th"
}

@article{Bianchi:2013finiteN,
      author         = "Bianchi, Marco S. and Leoni, Marta and Leoni, Matias and Mauri, Andrea and Penati, Silvia and Santambrogio, Alberto",
      title          = "{ABJM amplitudes and WL at finite $N$}",
      journal        = "JHEP",
      volume         = "09",
      year           = "2013",
      pages          = "114",
      doi            = "10.1007/JHEP09(2013)114",
      eprint         = "1306.3243",
      archivePrefix  = "arXiv",
      primaryClass   = "hep-th"
}

@article{Bianchi:2013nonplanar,
      author         = "Bianchi, Lorenzo and Bianchi, Marco S.",
      title          = "{Nonplanarity through unitarity in the ABJM theory}",
      journal        = "Phys. Rev. D",
      volume         = "89",
      year           = "2014",
      pages          = "125002",
      doi            = "10.1103/PhysRevD.89.125002",
      eprint         = "1311.6464",
      archivePrefix  = "arXiv",
      primaryClass   = "hep-th"
}

@article{Henn:2010abjmWilson,
      author         = "Henn, Johannes M. and Plefka, Jan and Wiegandt, Konstantin",
      title          = "{Light-like polygonal Wilson loops in 3d Chern--Simons and ABJM theory}",
      journal        = "JHEP",
      volume         = "08",
      year           = "2010",
      pages          = "032",
      doi            = "10.1007/JHEP08(2010)032",
      eprint         = "1004.0226",
      archivePrefix  = "arXiv",
      primaryClass   = "hep-th"
}

@article{Chen:2011abjmDualities,
      author         = "Chen, Wei-Ming and Huang, Yu-tin",
      title          = "{Dualities for Loop Amplitudes of $\mathcal{N}=6$ Chern--Simons Matter Theory}",
      journal        = "JHEP",
      volume         = "11",
      year           = "2011",
      pages          = "057",
      doi            = "10.1007/JHEP11(2011)057",
      eprint         = "1107.2710",
      archivePrefix  = "arXiv",
      primaryClass   = "hep-th"
}

@article{Bianchi:2011amplitudeWilson,
      author         = "Bianchi, Marco S. and Leoni, Matias and Mauri, Andrea and Penati, Silvia and Santambrogio, Alberto",
      title          = "{Scattering Amplitudes/Wilson Loop Duality in ABJM Theory}",
      journal        = "JHEP",
      volume         = "01",
      year           = "2012",
      pages          = "056",
      doi            = "10.1007/JHEP01(2012)056",
      eprint         = "1107.3139",
      archivePrefix  = "arXiv",
      primaryClass   = "hep-th"
}

@article{Bianchi:2011allorder,
      author         = "Bianchi, Marco S. and Leoni, Matias and Penati, Silvia",
      title          = "{An all order identity between ABJM and $\mathcal{N}=4$ SYM four-point amplitudes}",
      journal        = "JHEP",
      volume         = "04",
      year           = "2012",
      pages          = "045",
      doi            = "10.1007/JHEP04(2012)045",
      eprint         = "1112.3649",
      archivePrefix  = "arXiv",
      primaryClass   = "hep-th"
}

@article{CaronHuot:2012abjmSixPoint,
      author         = "Caron-Huot, Simon and Huang, Yu-tin",
      title          = "{The two-loop six-point amplitude in ABJM theory}",
      journal        = "JHEP",
      volume         = "03",
      year           = "2013",
      pages          = "075",
      doi            = "10.1007/JHEP03(2013)075",
      eprint         = "1210.4226",
      archivePrefix  = "arXiv",
      primaryClass   = "hep-th"
}

@article{Bianchi:2014threeLoopABJM,
      author         = "Bianchi, Marco S. and Leoni, Matias",
      title          = "{On the ABJM four-point amplitude at three loops and BDS exponentiation}",
      journal        = "JHEP",
      volume         = "11",
      year           = "2014",
      pages          = "077",
      doi            = "10.1007/JHEP11(2014)077",
      eprint         = "1403.3398",
      archivePrefix  = "arXiv",
      primaryClass   = "hep-th"
}

@article{Young:2014sia,
      author         = "Young, Donovan",
      title          = "{An Extremal Chiral Primary Three-Point Function at
                        Two-loops in ABJ(M)}",
      journal        = "JHEP",
      volume         = "12",
      year           = "2014",
      pages          = "141",
      doi            = "10.1007/JHEP12(2014)141",
      eprint         = "1411.0626",
      archivePrefix  = "arXiv",
      primaryClass   = "hep-th",
      reportNumber   = "QMUL-PH-14-23",
      SLACcitation   = "%%CITATION = ARXIV:1411.0626;%%"
}

@article{Bianchi:2011correlators,
      author         = "Bianchi, Marco S. and Leoni, Matias and Mauri, Andrea and Penati, Silvia and Ratti, CarloAlberto and Santambrogio, Alberto",
      title          = "{From Correlators to Wilson Loops in Chern--Simons Matter Theories}",
      journal        = "JHEP",
      volume         = "06",
      year           = "2011",
      pages          = "118",
      doi            = "10.1007/JHEP06(2011)118",
      eprint         = "1103.3675",
      archivePrefix  = "arXiv",
      primaryClass   = "hep-th"
}

@article{Siegel:1979wq,
	Author = {Siegel, Warren},
	Doi = {10.1016/0370-2693(79)90282-X},
	Journal = {Phys. Lett.},
	Pages = {193-196},
	Reportnumber = {HUTP-79/A006},
	Slaccitation = {%%CITATION = PHLTA,B84,193;%%},
	Title = {{Supersymmetric Dimensional Regularization via Dimensional Reduction}},
	Volume = {B84},
	Year = {1979}}

@article{Tkachov:1981wb,
	Author = {Tkachov, F. V.},
	Doi = {10.1016/0370-2693(81)90288-4},
	Journal = {Phys. Lett.},
	Pages = {65-68},
	Slaccitation = {%%CITATION = PHLTA,B100,65;%%},
	Title = {{A Theorem on Analytical Calculability of Four Loop Renormalization Group Functions}},
	Volume = {B100},
	Year = {1981}}

@article{Laporta:2001dd,
	Archiveprefix = {arXiv},
	Author = {Laporta, S.},
	Doi = {10.1142/S0217751X00002159},
	Eprint = {hep-ph/0102033},
	Journal = {Int. J. Mod. Phys.},
	Pages = {5087-5159},
	Primaryclass = {hep-ph},
	Slaccitation = {%%CITATION = HEP-PH/0102033;%%},
	Title = {{High precision calculation of multiloop Feynman integrals by difference equations}},
	Volume = {A15},
	Year = {2000}}

@article{Lee:1998bxa,
      author         = "Lee, Sangmin and Minwalla, Shiraz and Rangamani, Mukund
                        and Seiberg, Nathan",
      title          = "{Three point functions of chiral operators in D = 4, N=4
                        SYM at large N}",
      journal        = "Adv. Theor. Math. Phys.",
      volume         = "2",
      year           = "1998",
      pages          = "697-718",
      doi            = "10.4310/ATMP.1998.v2.n4.a1",
      eprint         = "hep-th/9806074",
      archivePrefix  = "arXiv",
      primaryClass   = "hep-th",
      reportNumber   = "PUPT-1796, IASSNS-HEP-98-51, PUPT-1796-",
      SLACcitation   = "%%CITATION = HEP-TH/9806074;%%"
}

@article{Eden:1999gh,
      author         = "Eden, B. and Howe, Paul S. and West, Peter C.",
      title          = "{Nilpotent invariants in N=4 SYM}",
      journal        = "Phys. Lett.",
      volume         = "B463",
      year           = "1999",
      pages          = "19-26",
      doi            = "10.1016/S0370-2693(99)00705-4",
      eprint         = "hep-th/9905085",
      archivePrefix  = "arXiv",
      primaryClass   = "hep-th",
      SLACcitation   = "%%CITATION = HEP-TH/9905085;%%"
}

@article{Heslop:2001gp,
      author         = "Heslop, P. J. and Howe, Paul S.",
      title          = "{OPEs and three-point correlators of protected operators
                        in N=4 SYM}",
      journal        = "Nucl. Phys.",
      volume         = "B626",
      year           = "2002",
      pages          = "265-286",
      doi            = "10.1016/S0550-3213(02)00023-8",
      eprint         = "hep-th/0107212",
      archivePrefix  = "arXiv",
      primaryClass   = "hep-th",
      SLACcitation   = "%%CITATION = HEP-TH/0107212;%%"
}

@article{Arutyunov:2001qw,
      author         = "Arutyunov, G. and Eden, B. and Sokatchev, E.",
      title          = "{On nonrenormalization and OPE in superconformal field
                        theories}",
      journal        = "Nucl. Phys.",
      volume         = "B619",
      year           = "2001",
      pages          = "359-372",
      doi            = "10.1016/S0550-3213(01)00529-6",
      eprint         = "hep-th/0105254",
      archivePrefix  = "arXiv",
      primaryClass   = "hep-th",
      reportNumber   = "AEI-2001-055, LAPTH-849-01",
      SLACcitation   = "%%CITATION = HEP-TH/0105254;%%"
}

@article{Chetyrkin:1981qh,
      author         = "Chetyrkin, K. G. and Tkachov, F. V.",
      title          = "{Integration by Parts: The Algorithm to Calculate beta
                        Functions in 4 Loops}",
      journal        = "Nucl. Phys.",
      volume         = "B192",
      year           = "1981",
      pages          = "159-204",
      doi            = "10.1016/0550-3213(81)90199-1",
      SLACcitation   = "%%CITATION = NUPHA,B192,159;%%"
}

@article{Kotikov:2003fb,
      author         = "Kotikov, A. V. and Lipatov, L. N. and Velizhanin, V. N.",
      title          = "{Anomalous dimensions of Wilson operators in N=4 SYM
                        theory}",
      journal        = "Phys. Lett.",
      volume         = "B557",
      year           = "2003",
      pages          = "114-120",
      doi            = "10.1016/S0370-2693(03)00184-9",
      eprint         = "hep-ph/0301021",
      archivePrefix  = "arXiv",
      primaryClass   = "hep-ph",
      SLACcitation   = "%%CITATION = HEP-PH/0301021;%%"
}

@article{Kotikov:2002ab,
      author         = "Kotikov, A. V. and Lipatov, L. N.",
      title          = "{DGLAP and BFKL evolution equations in the
                        $\mathcal{N}=4$ supersymmetric gauge theory}",
      journal        = "Nucl. Phys. B",
      volume         = "661",
      year           = "2003",
      pages          = "19-61",
      doi            = "10.1016/S0550-3213(03)00264-5",
      note           = "[Erratum: Nucl. Phys. B 685 (2004) 405--407]",
      eprint         = "hep-ph/0208220",
      archivePrefix  = "arXiv",
      primaryClass   = "hep-ph"
}

@article{Kotikov:2004er,
      author         = "Kotikov, A. V. and Lipatov, L. N. and Onishchenko, A. I.
                        and Velizhanin, V. N.",
      title          = "{Three loop universal anomalous dimension of the Wilson
                        operators in $N=4$ SUSY Yang-Mills model}",
      journal        = "Phys. Lett.",
      volume         = "B595",
      year           = "2004",
      pages          = "521-529",
      doi            = "10.1016/j.physletb.2004.05.078",
      note           = "[Erratum: Phys. Lett.B632,754(2006)]",
      eprint         = "hep-th/0404092",
      archivePrefix  = "arXiv",
      primaryClass   = "hep-th",
      reportNumber   = "WSU-HEP-0405",
      SLACcitation   = "%%CITATION = HEP-TH/0404092;%%"
}

@article{Beisert:2006ez,
      author         = "Beisert, Niklas and Eden, Burkhard and Staudacher,
                        Matthias",
      title          = "{Transcendentality and Crossing}",
      journal        = "J. Stat. Mech.",
      volume         = "0701",
      year           = "2007",
      pages          = "P01021",
      doi            = "10.1088/1742-5468/2007/01/P01021",
      eprint         = "hep-th/0610251",
      archivePrefix  = "arXiv",
      primaryClass   = "hep-th",
      reportNumber   = "AEI-2006-079, ITP-UU-06-44, SPIN-06-34",
      SLACcitation   = "%%CITATION = HEP-TH/0610251;%%"
}

@article{Eden:2006rx,
      author         = "Eden, Burkhard and Staudacher, Matthias",
      title          = "{Integrability and transcendentality}",
      journal        = "J. Stat. Mech.",
      volume         = "0611",
      year           = "2006",
      pages          = "P11014",
      doi            = "10.1088/1742-5468/2006/11/P11014",
      eprint         = "hep-th/0603157",
      archivePrefix  = "arXiv",
      primaryClass   = "hep-th",
      reportNumber   = "AEI-2005-165",
      SLACcitation   = "%%CITATION = HEP-TH/0603157;%%"
}

@article{Bianchi:2020cfn,
    author = "Bianchi, Marco S.",
    title = "{On three-point functions in ABJM and the latitude Wilson loop}",
    eprint = "2005.09522",
    archivePrefix = "arXiv",
    primaryClass = "hep-th",
    doi = "10.1007/JHEP10(2020)075",
    journal = "JHEP",
    volume = "10",
    pages = "075",
    year = "2020"
}

@article{Vermaseren:2000nd,
    author = "Vermaseren, J. A. M.",
    title = "{New features of FORM}",
    eprint = "math-ph/0010025",
    archivePrefix = "arXiv",
    month = "10",
    year = "2000"
}

@article{Ruijl:2017dtg,
    author = "Ruijl, Ben and Ueda, Takahiro and Vermaseren, Jos",
    title = "{FORM version 4.2}",
    eprint = "1707.06453",
    archivePrefix = "arXiv",
    primaryClass = "hep-ph",
    month = "7",
    year = "2017"
}

@article{Ruijl:2017cxj,
    author = "Ruijl, B. and Ueda, T. and Vermaseren, J. A. M.",
    title = "{Forcer, a FORM program for the parametric reduction of four-loop massless propagator diagrams}",
    eprint = "1704.06650",
    archivePrefix = "arXiv",
    primaryClass = "hep-ph",
    reportNumber = "NIKHEF-2017-019",
    doi = "10.1016/j.cpc.2020.107198",
    journal = "Comput. Phys. Commun.",
    volume = "253",
    pages = "107198",
    year = "2020"
}

@article{vanRitbergen:1998pn,
    author = "van Ritbergen, T. and Schellekens, A. N. and Vermaseren, J. A. M.",
    title = "{Group theory factors for Feynman diagrams}",
    eprint = "hep-ph/9802376",
    archivePrefix = "arXiv",
    reportNumber = "UM-TH-98-01, NIKHEF-98-004",
    doi = "10.1142/S0217751X99000038",
    journal = "Int. J. Mod. Phys. A",
    volume = "14",
    pages = "41--96",
    year = "1999"}

@article{Nogueira:1991ex,
    author = "Nogueira, Paulo",
    title = "{Automatic Feynman graph generation}",
    reportNumber = "IFM-7-91",
    doi = "10.1006/jcph.1993.1074",
    journal = "J. Comput. Phys.",
    volume = "105",
    pages = "279--289",
    year = "1993"
}

@article{Lee:2011jt,
    author = "Lee, R. N. and Smirnov, A. V. and Smirnov, V. A.",
    title = "{Master Integrals for Four-Loop Massless Propagators up to Transcendentality Weight Twelve}",
    eprint = "1108.0732",
    archivePrefix = "arXiv",
    primaryClass = "hep-th",
    reportNumber = "TTP11-20, SFB-CPP-11-38",
    doi = "10.1016/j.nuclphysb.2011.11.005",
    journal = "Nucl. Phys. B",
    volume = "856",
    pages = "95--110",
    year = "2012"
}

@article{Intriligator:1999ff,
    author = "Intriligator, Kenneth A. and Skiba, Witold",
    title = "{Bonus symmetry and the operator product expansion of N=4 SuperYang-Mills}",
    eprint = "hep-th/9905020",
    archivePrefix = "arXiv",
    reportNumber = "UCSD-PTH-99-06, IASSNS-HEP-99-45",
    doi = "10.1016/S0550-3213(99)00430-7",
    journal = "Nucl. Phys. B",
    volume = "559",
    pages = "165--183",
    year = "1999"
}

@article{Henn:2013pwa,
    author = "Henn, Johannes M.",
    title = "{Multiloop integrals in dimensional regularization made simple}",
    eprint = "1304.1806",
    archivePrefix = "arXiv",
    primaryClass = "hep-th",
    doi = "10.1103/PhysRevLett.110.251601",
    journal = "Phys. Rev. Lett.",
    volume = "110",
    pages = "251601",
    year = "2013"
}

@article{Baikov:2010hf,
    author = "Baikov, P. A. and Chetyrkin, K. G.",
    title = "{Four Loop Massless Propagators: An Algebraic Evaluation of All Master Integrals}",
    eprint = "1004.1153",
    archivePrefix = "arXiv",
    primaryClass = "hep-ph",
    reportNumber = "TTP10-18, SFB-CPP-10-24",
    doi = "10.1016/j.nuclphysb.2010.05.004",
    journal = "Nucl. Phys. B",
    volume = "837",
    pages = "186--220",
    year = "2010"
}

@article{vanNeerven:1985ja,
    author = "van Neerven, W. L.",
    title = "{Infrared Behavior of On-shell Form-factors in a $N=4$ Supersymmetric {Yang-Mills} Field Theory}",
    reportNumber = "DO-TH-85-19",
    doi = "10.1007/BF01571808",
    journal = "Z. Phys. C",
    volume = "30",
    pages = "595",
    year = "1986"
}

@article{Gehrmann:2011xn,
    author = "Gehrmann, Thomas and Henn, Johannes M. and Huber, Tobias",
    title = "{The three-loop form factor in N=4 super Yang-Mills}",
    eprint = "1112.4524",
    archivePrefix = "arXiv",
    primaryClass = "hep-th",
    reportNumber = "HU-EP-11-11-61, NSF-KITP-11-268, ZU-TH-28-11, SI-HEP-2011-19",
    doi = "10.1007/JHEP03(2012)101",
    journal = "JHEP",
    volume = "03",
    pages = "101",
    year = "2012"
}

@article{Huber:2019fxe,
    author = "Huber, Tobias and von Manteuffel, Andreas and Panzer, Erik and Schabinger, Robert M. and Yang, Gang",
    title = "{The four-loop cusp anomalous dimension from the $N=4$ Sudakov form factor}",
    eprint = "1912.13459",
    archivePrefix = "arXiv",
    primaryClass = "hep-th",
    reportNumber = "MSUHEP-19-030, P3H-19-061",
    doi = "10.1016/j.physletb.2020.135543",
    journal = "Phys. Lett. B",
    volume = "807",
    pages = "135543",
    year = "2020"
}

@article{Agarwal:2021zft,
    author = "Agarwal, Bakul and von Manteuffel, Andreas and Panzer, Erik and Schabinger, Robert M.",
    title = "{Four-loop collinear anomalous dimensions in QCD and N=4 super Yang-Mills}",
    eprint = "2102.09725",
    archivePrefix = "arXiv",
    primaryClass = "hep-ph",
    reportNumber = "MSUHEP-21-006",
    doi = "10.1016/j.physletb.2021.136503",
    journal = "Phys. Lett. B",
    volume = "820",
    pages = "136503",
    year = "2021"
}

@article{Lee:2021lkc,
    author = "Lee, Roman N. and von Manteuffel, Andreas and Schabinger, Robert M. and Smirnov, Alexander V. and Smirnov, Vladimir A. and Steinhauser, Matthias",
    title = "{The four-loop $ \mathcal{N} $ = 4 SYM Sudakov form factor}",
    eprint = "2110.13166",
    archivePrefix = "arXiv",
    primaryClass = "hep-th",
    reportNumber = "MSUHEP-21-029, P3H-21-082, TTP21-043",
    doi = "10.1007/JHEP01(2022)091",
    journal = "JHEP",
    volume = "01",
    pages = "091",
    year = "2022"
}

@article{Bork:2010wf,
    author = "Bork, L. V. and Kazakov, D. I. and Vartanov, G. S.",
    title = "{On form factors in N=4 sym}",
    eprint = "1011.2440",
    archivePrefix = "arXiv",
    primaryClass = "hep-th",
    doi = "10.1007/JHEP02(2011)063",
    journal = "JHEP",
    volume = "02",
    pages = "063",
    year = "2011"
}

@article{Brandhuber:2014ica,
    author = "Brandhuber, Andreas and Penante, Brenda and Travaglini, Gabriele and Wen, Congkao",
    title = "{The last of the simple remainders}",
    eprint = "1406.1443",
    archivePrefix = "arXiv",
    primaryClass = "hep-th",
    reportNumber = "QMUL-PH-14-09",
    doi = "10.1007/JHEP08(2014)100",
    journal = "JHEP",
    volume = "08",
    pages = "100",
    year = "2014"
}

@article{Banerjee:2016kri,
    author = "Banerjee, Pulak and Dhani, Prasanna K. and Mahakhud, Maguni and Ravindran, V. and Seth, Satyajit",
    title = "{Finite remainders of the Konishi at two loops in $ \mathcal{N}=4 $ SYM}",
    eprint = "1612.00885",
    archivePrefix = "arXiv",
    primaryClass = "hep-th",
    doi = "10.1007/JHEP05(2017)085",
    journal = "JHEP",
    volume = "05",
    pages = "085",
    year = "2017"
}

@article{Lin:2020dyj,
    author = "Lin, Guanda and Yang, Gang",
    title = "{Non-planar form factors of generic local operators via on-shell unitarity and color-kinematics duality}",
    eprint = "2011.06540",
    archivePrefix = "arXiv",
    primaryClass = "hep-th",
    doi = "10.1007/JHEP04(2021)176",
    journal = "JHEP",
    volume = "04",
    pages = "176",
    year = "2021"
}

@article{Brandhuber:2012vm,
    author = "Brandhuber, Andreas and Travaglini, Gabriele and Yang, Gang",
    title = "{Analytic two-loop form factors in N=4 SYM}",
    eprint = "1201.4170",
    archivePrefix = "arXiv",
    primaryClass = "hep-th",
    reportNumber = "QMUL-PH-12-01, WIS-03-12-JAN-DPPA",
    doi = "10.1007/JHEP05(2012)082",
    journal = "JHEP",
    volume = "05",
    pages = "082",
    year = "2012"
}

@article{Bern:2005iz,
    author = "Bern, Zvi and Dixon, Lance J. and Smirnov, Vladimir A.",
    title = "{Iteration of planar amplitudes in maximally supersymmetric Yang-Mills theory at three loops and beyond}",
    eprint = "hep-th/0505205",
    archivePrefix = "arXiv",
    reportNumber = "SLAC-PUB-11210, UCLA-05-TEP-14",
    doi = "10.1103/PhysRevD.72.085001",
    journal = "Phys. Rev. D",
    volume = "72",
    pages = "085001",
    year = "2005"
}

@article{DelDuca:2009au,
    author = "Del Duca, Vittorio and Duhr, Claude and Smirnov, Vladimir A.",
    title = "{An Analytic Result for the Two-Loop Hexagon Wilson Loop in N = 4 SYM}",
    eprint = "0911.5332",
    archivePrefix = "arXiv",
    primaryClass = "hep-ph",
    reportNumber = "IPPP-09-92, DCPT-09-184",
    doi = "10.1007/JHEP03(2010)099",
    journal = "JHEP",
    volume = "03",
    pages = "099",
    year = "2010"
}

@article{DelDuca:2010zg,
    author = "Del Duca, Vittorio and Duhr, Claude and Smirnov, Vladimir A.",
    title = "{The Two-Loop Hexagon Wilson Loop in N = 4 SYM}",
    eprint = "1003.1702",
    archivePrefix = "arXiv",
    primaryClass = "hep-th",
    reportNumber = "IPPP-10-21, DCPT-10-42, CERN-PH-TH-2010-059",
    doi = "10.1007/JHEP05(2010)084",
    journal = "JHEP",
    volume = "05",
    pages = "084",
    year = "2010"
}

@article{Goncharov:2010jf,
    author = "Goncharov, Alexander B. and Spradlin, Marcus and Vergu, C. and Volovich, Anastasia",
    title = "{Classical Polylogarithms for Amplitudes and Wilson Loops}",
    eprint = "1006.5703",
    archivePrefix = "arXiv",
    primaryClass = "hep-th",
    reportNumber = "BROWN-HET-1602",
    doi = "10.1103/PhysRevLett.105.151605",
    journal = "Phys. Rev. Lett.",
    volume = "105",
    pages = "151605",
    year = "2010"
}

@article{Davies:2026cci,
    author = "Davies, J. and Kaneko, T. and Marinissen, C. and Ueda, T. and Vermaseren, J. A. M.",
    title = "{FORM Version 5.0}",
    eprint = "2601.19982",
    archivePrefix = "arXiv",
    primaryClass = "hep-ph",
    month = "1",
    year = "2026"
}

@article{Bianchi:2023llc,
    author = "Bianchi, Marco S.",
    title = "{Protected and uniformly transcendental}",
    eprint = "2306.06239",
    archivePrefix = "arXiv",
    primaryClass = "hep-th",
    doi = "10.1007/JHEP09(2023)121",
    journal = "JHEP",
    volume = "09",
    pages = "121",
    year = "2023"
}

@article{Bianchi:2025sjc,
    author = "Bianchi, Marco S.",
    title = "{Uniformly transcendental bases for protected two-point functions}",
    eprint = "2512.11516",
    archivePrefix = "arXiv",
    primaryClass = "hep-th",
    doi = "10.1007/JHEP04(2026)207",
    journal = "JHEP",
    volume = "04",
    pages = "207",
    year = "2026"
}

@article{Chakraborty:2026hdv,
    author = "Chakraborty, Mrigankamauli and Moch, Sven-Olaf",
    title = "{Dimensional Reduction is Supersymmetric at Three Loops}",
    eprint = "2603.02892",
    archivePrefix = "arXiv",
    primaryClass = "hep-th",
    reportNumber = "PUBDB-2026-00811",
    month = "3",
    year = "2026"
}

@article{Bianchi:2024nah,
    author = "Bianchi, Marco S.",
    title = "{Transcendentality of ABJM two-point functions}",
    eprint = "2410.23395",
    archivePrefix = "arXiv",
    primaryClass = "hep-th",
    doi = "10.1007/JHEP12(2024)188",
    journal = "JHEP",
    volume = "12",
    pages = "188",
    year = "2024"
}

@article{Dey:2011ea,
    author = "Dey, Tanay K.",
    title = "{Exact Large R-charge Correlators in ABJM Theory}",
    eprint = "1105.0218",
    archivePrefix = "arXiv",
    primaryClass = "hep-th",
    doi = "10.1007/JHEP08(2011)066",
    journal = "JHEP",
    volume = "08",
    pages = "066",
    year = "2011"
}

@article{Chakrabortty:2011gv,
    author = "Chakrabortty, Shankhadeep and Dey, Tanay K.",
    title = "{Correlators of Giant Gravitons from dual ABJ(M) Theory}",
    eprint = "1112.6299",
    archivePrefix = "arXiv",
    primaryClass = "hep-th",
    doi = "10.1007/JHEP03(2012)062",
    journal = "JHEP",
    volume = "03",
    pages = "062",
    year = "2012"
}

@article{Caputa:2012pi,
    author = "Caputa, Pawel and Mohammed, Badr Awad Elseid",
    title = "{From Schurs to Giants in ABJ(M)}",
    eprint = "1210.7705",
    archivePrefix = "arXiv",
    primaryClass = "hep-th",
    doi = "10.1007/JHEP01(2013)055",
    journal = "JHEP",
    volume = "01",
    pages = "055",
    year = "2013"
}

@article{Georgoudis:2026loc,
    author = "Georgoudis, Alessandro and Minahan, Joseph A. and Nedelin, Anton and Wen, Congkao",
    title = "{Two-point functions in $4-2\epsilon$ dimensions from localization}",
    eprint = "2606.18442",
    archivePrefix = "arXiv",
    primaryClass = "hep-th",
    month = "6",
    year = "2026"
}

@article{Aharony:2008ug,
    author = "Aharony, Ofer and Bergman, Oren and Jafferis, Daniel Louis and Maldacena, Juan",
    title = "{N=6 superconformal Chern-Simons-matter theories, M2-branes and their gravity duals}",
    eprint = "0806.1218",
    archivePrefix = "arXiv",
    primaryClass = "hep-th",
    doi = "10.1088/1126-6708/2008/10/091",
    journal = "JHEP",
    volume = "10",
    pages = "091",
    year = "2008"
}

@article{Aharony:2008gk,
    author = "Aharony, Ofer and Bergman, Oren and Jafferis, Daniel Louis",
    title = "{Fractional M2-branes}",
    eprint = "0807.4924",
    archivePrefix = "arXiv",
    primaryClass = "hep-th",
    doi = "10.1088/1126-6708/2008/11/043",
    journal = "JHEP",
    volume = "11",
    pages = "043",
    year = "2008"
}

@article{Kapustin:2009kz,
    author = "Kapustin, Anton and Willett, Brian and Yaakov, Itamar",
    title = "{Exact Results for Wilson Loops in Superconformal Chern--Simons Theories with Matter}",
    eprint = "0909.4559",
    archivePrefix = "arXiv",
    primaryClass = "hep-th",
    doi = "10.1007/JHEP03(2010)089",
    journal = "JHEP",
    volume = "03",
    pages = "089",
    year = "2010"
}

@article{Dedushenko:2016jxl,
    author = "Dedushenko, Mykola and Pufu, Silviu S. and Yacoby, Ran",
    title = "{A one-dimensional theory for Higgs branch operators}",
    eprint = "1610.00740",
    archivePrefix = "arXiv",
    primaryClass = "hep-th",
    doi = "10.1007/JHEP03(2018)138",
    journal = "JHEP",
    volume = "03",
    pages = "138",
    year = "2018"
}

@article{Gorini:2020new,
    author = "Gorini, Nicola and Griguolo, Luca and Guerrini, Luigi and Penati, Silvia and Seminara, Domenico and Soresina, Paolo",
    title = "{The topological line of ABJ(M) theory}",
    eprint = "2012.11613",
    archivePrefix = "arXiv",
    primaryClass = "hep-th",
    doi = "10.1007/JHEP06(2021)091",
    journal = "JHEP",
    volume = "06",
    pages = "091",
    year = "2021"
}

@article{Guerrini:2021zuk,
    author = "Guerrini, Luigi and Penati, Silvia and Yaakov, Itamar",
    title = "{Generating functions for Higgs/Coulomb branch operators from 1d--3d cohomological equivalence}",
    eprint = "2112.13816",
    archivePrefix = "arXiv",
    primaryClass = "hep-th",
    doi = "10.1007/JHEP04(2022)171",
    journal = "JHEP",
    volume = "04",
    pages = "171",
    year = "2022"
}

@article{Armanini:2024loc,
    author = "Armanini, Elisabetta and Griguolo, Luca and Guerrini, Luigi",
    title = "{BPS Wilson loops in mass-deformed ABJM theory: Fermi gas expansions and new defect CFT data}",
    eprint = "2401.12288",
    archivePrefix = "arXiv",
    primaryClass = "hep-th",
    doi = "10.21468/SciPostPhys.17.2.035",
    journal = "SciPost Phys.",
    volume = "17",
    number = "2",
    pages = "035",
    year = "2024"
}

@article{Baggio:2012rr,
    author = "Baggio, Marco and de Boer, Jan and Papadodimas, Kyriakos",
    title = "{A non-renormalization theorem for chiral primary 3-point functions}",
    eprint = "1203.1036",
    archivePrefix = "arXiv",
    primaryClass = "hep-th",
    doi = "10.1007/JHEP07(2012)137",
    journal = "JHEP",
    volume = "07",
    pages = "137",
    year = "2012"
}

@article{Bianchi:2026tracing,
    author = "Bianchi, Marco S.",
    title = "{Tracing Transcendentality in Protected Correlators of N=4 SYM}",
    eprint = "2606.18362",
    archivePrefix = "arXiv",
    primaryClass = "hep-th",
    month = "6",
    year = "2026"
}

@article{Lee:2015summertime,
    author = "Lee, Roman N. and Mingulov, Kirill T.",
    title = "{Introducing SummerTime: a package for high-precision computation of sums appearing in DRA method}",
    eprint = "1507.04256",
    archivePrefix = "arXiv",
    primaryClass = "hep-ph",
    doi = "10.1016/j.cpc.2016.02.018",
    journal = "Comput. Phys. Commun.",
    volume = "203",
    pages = "255--267",
    year = "2016"
}

@article{Lee:2010yangian,
    author = "Lee, Sangmin",
    title = "{Yangian Invariant Scattering Amplitudes in Supersymmetric Chern--Simons Theory}",
    eprint = "1007.4772",
    archivePrefix = "arXiv",
    primaryClass = "hep-th",
    doi = "10.1103/PhysRevLett.105.151603",
    journal = "Phys. Rev. Lett.",
    volume = "105",
    pages = "151603",
    year = "2010"
}

@article{Huang:2013orthogonal,
    author = "Huang, Yu-tin and Wen, Congkao",
    title = "{ABJM amplitudes and the positive orthogonal Grassmannian}",
    eprint = "1309.3252",
    archivePrefix = "arXiv",
    primaryClass = "hep-th",
    doi = "10.1007/JHEP02(2014)104",
    journal = "JHEP",
    volume = "02",
    pages = "104",
    year = "2014"
}

@article{Huang:2014loopGrassmannian,
    author = "Huang, Yu-tin and Wen, Congkao and Xie, Dan",
    title = "{The Positive Orthogonal Grassmannian and Loop Amplitudes of ABJM}",
    eprint = "1402.1479",
    archivePrefix = "arXiv",
    primaryClass = "hep-th",
    doi = "10.1088/1751-8113/47/47/474008",
    journal = "J. Phys. A",
    volume = "47",
    pages = "474008",
    year = "2014"
}

@article{Corley:2001zk,
    author = "Corley, Steve and Jevicki, Antal and Ramgoolam, Sanjaye",
    title = "{Exact Correlators of Giant Gravitons from dual N=4 SYM Theory}",
    eprint = "hep-th/0111222",
    archivePrefix = "arXiv",
    doi = "10.4310/ATMP.2001.v5.n4.a6",
    journal = "Adv. Theor. Math. Phys.",
    volume = "5",
    pages = "809--839",
    year = "2002"
}

@article{DeComarmond:2010spectra,
    author = "De Comarmond, Vincent and de Mello Koch, Robert and Jefferies, Katherine",
    title = "{Surprisingly Simple Spectra}",
    eprint = "1012.3884",
    archivePrefix = "arXiv",
    primaryClass = "hep-th",
    doi = "10.1007/JHEP02(2011)006",
    journal = "JHEP",
    volume = "02",
    pages = "006",
    year = "2011"
}

@article{Brown:2007multimatrix,
    author = "Brown, T. W. and Heslop, P. J. and Ramgoolam, S.",
    title = "{Diagonal Multi-Matrix Correlators and BPS Operators in N=4 SYM}",
    eprint = "0711.0176",
    archivePrefix = "arXiv",
    primaryClass = "hep-th",
    doi = "10.1088/1126-6708/2008/02/030",
    journal = "JHEP",
    volume = "02",
    pages = "030",
    year = "2008"
}

@article{Bhattacharyya:2008multimatrix,
    author = "Bhattacharyya, Rajsekhar and Collins, Storm and de Mello Koch, Robert",
    title = "{Exact Multi-Matrix Correlators}",
    eprint = "0801.2061",
    archivePrefix = "arXiv",
    primaryClass = "hep-th",
    doi = "10.1088/1126-6708/2008/03/044",
    journal = "JHEP",
    volume = "03",
    pages = "044",
    year = "2008"
}

@article{Brown:2008permutations,
    author = "Brown, T. W.",
    title = "{Permutations and the Loop}",
    eprint = "0801.2094",
    archivePrefix = "arXiv",
    primaryClass = "hep-th",
    doi = "10.1088/1126-6708/2008/06/008",
    journal = "JHEP",
    volume = "06",
    pages = "008",
    year = "2008"
}

\end{document}